\documentclass[a4paper,12pt]{article}

\usepackage[a4paper,margin=2cm, marginparwidth=2cm]{geometry}

\usepackage{amsmath,amssymb}

\usepackage{graphicx}
\graphicspath{{figures/}}

\usepackage{hyperref}

\newcommand{\bea}{\begin{eqnarray}}
\newcommand{\eea}{\end{eqnarray}}
\newcommand{\beq}{\begin{equation}}
\newcommand{\eeq}{\end{equation}}

\providecommand{\eqref}[1]{equation (\ref{#1})}
\newcommand{\figref}[1]{Fig.\ \ref{#1}}
\newcommand{\Figref}[1]{Fig.\ \ref{#1}} 
\newcommand{\secref}[1]{Section \ref{#1}}
\newcommand{\tabref}[1]{Table \ref{#1}}

\newcommand{\appref}[2]{Appendix {#2} of the Supplementary Information}

\newcommand{\tnote}[1]{} 
\newcommand{\tcomment}[1]{} 

\providecommand{\href}[2]{{#2}\footnote{See \texttt{#1}}}

\newcommand{\Tcal}{\mathcal{T}}

\newcommand{\Fmin}{F_\mathrm{min}}
\newcommand{\Fhat}{\widehat{F}}

\newcommand{\searchparam}{m}
\newcommand{\oppparam}{n}

\newcommand{\flowparam}{t}
\newcommand{\outflowvalue}{O}
\newcommand{\inflowvalue}{I}
\newcommand{\popvalue}{P}

\begin{document}



\renewcommand{\thefootnote}{\fnsymbol{footnote}}

\begin{flushright}
\href{https://arxiv.org/abs/1909.07194}{\texttt{arXiv:1909.07194}} \\
Scientific Reports \textbf{10} (2020) 17474 \\
\href{http://doi.org/10.1038/s41598-020-74601-z.}{DOI: \texttt{10.1038/s41598-020-74601-z}}
\end{flushright}
\vspace*{0.5cm}

\begin{center}
{\LARGE\textbf{Predictive limitations of spatial interaction models: a non-Gaussian analysis}}
\\[6pt]
 {\large
 B.\ Hilton\textsuperscript{*},
 A.\ P.\ Sood\textsuperscript{*},
 \href{http://www.imperial.ac.uk/people/t.evans}{T.\ S.\ Evans},
 }
\\[0.5\baselineskip]
(*) Equal first authors 
\\[0.5\baselineskip]
\href{https://www.imperial.ac.uk/complexity-science}{Centre for Complexity Science}, and \href{http://www.imperial.ac.uk/theoreticalphysics}{Theoretical Physics Group},\\
Physics Dept., Imperial College London, SW7 2AZ, U.K.
\\[0.5\baselineskip]
10th September 2020
\end{center}

\begin{abstract}
We present a method to compare spatial interaction models against data based on well known statistical measures that are appropriate for such models and data. We illustrate our approach using a widely used example: commuting data, specifically from the US Census 2000. We find that the radiation model performs significantly worse than an appropriately chosen simple gravity model. Various conclusions are made regarding the development and use of spatial interaction models, including: that spatial interaction models fit badly to data in an absolute sense, that therefore the risk of over-fitting is small and adding additional fitted parameters improves the predictive power of models, and that appropriate choices of input data can improve model fit.
\end{abstract}

\section{Introduction}

The ability to predict the number of vehicles, the amount of goods, or the spread of disease between two locations, using only limited data about each location, is important in a variety of academic disciplines. Problems of this nature can be studied using `spatial interaction models'. Given some measures of the importance of each site $i$,
and the distance $d_{ij}$ between two sites $i$ and $j$, these models predict the flow from site $i$ to site $j$, denoted $F_{ij}$.
The distance $d_{ij}$ need not be a geographical distance; it could reflect the cost of travel or other socio-economic measures of separation. These models only predict flows between distinct sites, and so $i \neq j$.

The nature of spatial interaction models and the associated data means that residual errors cannot always be assumed to be Gaussian,  though this is often assumed in the literature. Our primary goal is to improve upon the statistical analysis commonly carried out in the literature and apply this improved analysis to determine the relative effectiveness of key examples from two popular families of models: gravity models and radiation models. Additionally, our methods are used to identify which features of these models give the greatest improvement in results.

We will start by reviewing the data used in our work. In \secref{s:models}, we will look at the various spatial interaction models we consider. The statistical methods used are described in \secref{s:stats} with more details on alternatives used in the literature given in \appref{a:common}{D}.  Our results are then shown in \secref{s:results}. We will conclude with a discussion of our work. A summary of the notation used in this paper is provided in \appref{a:notation}{A}.

\section{Data}\label{s:datasets}


It is inherent to the nature of statistical analysis that models must be compared against data. In this paper we wish to focus on the features of spatial models and on the features of different analysis methods used to study spatial data and models.
To do this we sought a dataset which acts as a standard to be used when comparing different models and different analysis techniques.
It is essential then that such a standard is an open data set and it would be useful if the standard dataset was already well known and well studied to give authors many sources of independent information on the standard dataset.
We have chosen to work with the US Census 2000: the county-to-county worker flow data from the US Census 2000\cite{US_census_2000}.
It is both an open source dataset and widely used.

In particular, the US Census 2000 datset was used in Simini et al.\cite{SGMB12} when developing the Radiation model.
This ensures that any differences between our results and those of Simini et al.\cite{SGMB12} arise due to changes in the analysis rather than simply the choice of data. Using this data, the radiation model was compared favourably against the gravity model\cite{SGMB12}.

As a further check and to verify that our conclusions are a result of the models and the US commuter flow system rather than merely a feature of the specific data set, we also used the parallel data set from the American Community Survey\cite{ACS_2013} 2009--2013. We obtained the populations of the counties at the census dates of 2000 \cite{cest_1990} and 2010 \cite{cest_2000}. Though we often use the language of commuting to describe our approach, our methods are data set agnostic, and therefore our results have wider applicability.

\begin{figure}[htb!]
 	\centering
 	\includegraphics[width=0.8\textwidth]{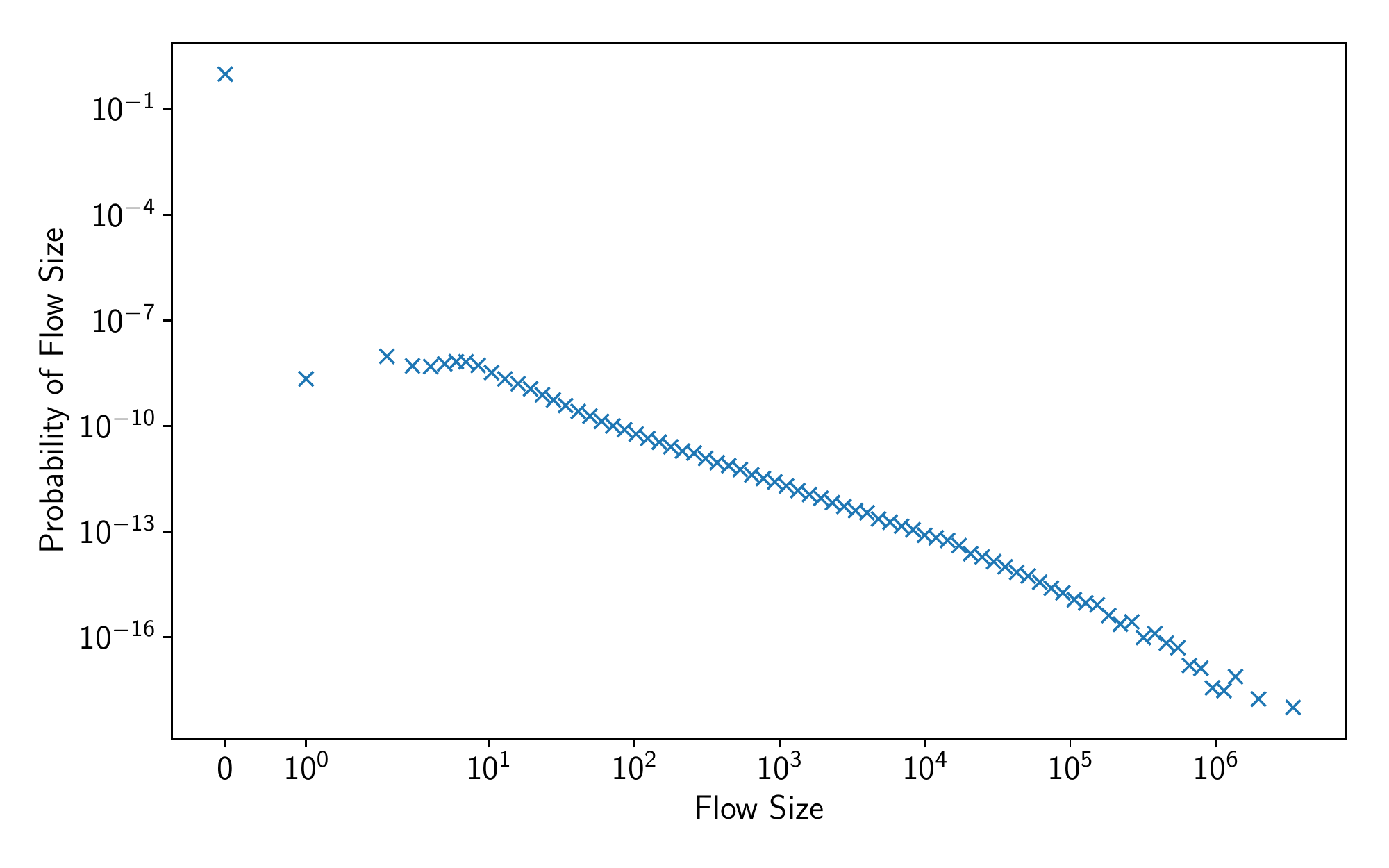} 
 	\caption{The distribution of commuter flow sizes in the US Census 2000 data\cite{US_census_2000}. }
 	\label{fig:censusflow}
\end{figure}

In the US Census 2000, there are 3109 counties or their equivalents within the 48 contiguous United States. These form the sites used by our spatial models. The US Census 2000 asked, for each person listed: ``at what location did this person work \textit{last week}?'' Respondents were further instructed ``if this person worked at more than one location, [to] print where he or she worked most last week.'' This means that our figures for commuting will include data from those who occasionally work at other locations for a few days and these are likely to inflate the number of long distance trips recorded relative to data representing where a person worked for most of a year. Information on the distribution of flows is shown in \figref{fig:censusflow} and \appref{a:data}{E}.

From this data we define three values associated with each site $i$, which are generic to many spatial interaction contexts: the site population $P_i$, the flow into a site $I_i$, and the flow out  $O_i$. While these three values are likely to be correlated at each site for our commuting data, there are large individual differences as sites may have developed specialised functions.  For instance in the US Census 2002 data\cite{US_census_2000}, many people work in San Francisco county who commute in from other counties (265,291 people), but fewer live in San Francisco county and commute elsewhere (130,036 people).

We use this data on the population and the number of commuters arriving and leaving a site to determine model parameters associated with site importance. We use $w_i$ (site weight) as a generic site importance model parameter but, depending on the model, we can use up to three more specific site parameters to characterise a site: a repulsiveness parameter $t_i$ controlling the total flow out of a site, an attractiveness parameter $n_i$ that controls the flow into a site, and in some cases an `aspiration' parameter $m_i$ that controls how far a commuter will travel.

The distances needed for the models were great-circle distances between the geographical centres of each pair of US counties. These data were obtained from the National Bureau of Economic Research\cite{distance_data}.

\section{Models} \label{s:models}

\subsection{Gravity Models}\label{sec:gravmodel}

One of the most widely used spatial interaction models is a class of models known as `gravity models', which have been used in a variety of socio-economic contexts since the 19th Century but have seen much development since the 1950s (see elsewhere\cite{ES90,NR12} for general reviews).

The simplest gravity model is given by
\begin{equation}
 \Fhat_{ij}  =   w_i w_j f(d_{ij})  \label{e:gravitysimple}
 \, ,
\end{equation}
where $\Fhat_{ij}$ is the model's estimate of the flow $F_{ij}$ from $i$ to $j$. The $w_i$ and $w_j$ parameters are the weights of sites $i$ and $j$ respectively, some measure of the importance of sites. The function $f(d_{ij})$ is some monotonically decreasing function of (generalised) distance: the `deterrence function'. This function is often chosen without theoretical motivation and typically includes additional parameters; these must be determined using previously known data.
Such flexibility in the form of the deterrence function can be regarded as a key limitation of the gravity model\cite{SGMB12}. However, in practice simple forms are often found to be effective. Common deterrence functions include exponentials\cite{balcan_multiscale_2009} ($f(x) = e^{-\beta x}$ for some $\beta > 0$) and power laws\cite{kaluza_complex_2010,viboud_synchrony_2006} ($f(x) = x^{-\beta}$ for some $\beta > 0$). The deterrence functions invariably include a global parameter, $\beta$ in our examples, that is the same for all pairs of sites. This might be set from data, for instance $\beta^{-1}$ represents a typical length scale for the exponential form. However, such global model parameters are often determined by varying their values until the model has the best possible fit to the data.

In order to accurately test the extent of the difference in predictive power between models, they must share any feature that is not being explicitly compared. All the models considered here are `production constrained' models in which the output of each site is fixed by a model parameter for that site. So rather than the simplest gravity model of \eqref{e:gravitysimple}, we will use a production constrained gravity model\cite{ES90,W67,W71} of the form
\begin{equation}
    \Fhat_{ij} = \frac{ t_i n_j d_{ij}^{-\beta}}{\sum_k n_k d_{ik}^{-\beta}}
      \, .
      \quad (i \neq j)
    \label{e:gravitypc}
\end{equation}
This obeys $\sum_j \Fhat_{ij} = t_i$, the production constraint making the site model parameter $t_i$ equal to the total flow leaving site $i$. The $n_j$ parameter is some measure of the `attractiveness' of site $j$ that controls the flow into each site, though this is not necessarily equal to the flow into site $j$.
Even if $t_i=n_i$ (as is often assumed), it is worth noting that this model already describes an asymmetric flow with $\Fhat_{ij}\neq\Fhat_{ji}$ in general. Thus, unlike the simple gravity model, this production constrained gravity model can produce flow asymmetries akin to those that are present in real data, as illustrated by the example of San Francisco county considered in \secref{s:datasets}.

For our work with the gravity model \eqref{e:gravitypc}, we will set the output site parameter equal to the number of commuters leaving a site, $t_i=O_i$, while the site attractiveness parameter will be set equal to the number of commuters arriving $n_i=I_i$. We will choose the single global model parameter $\beta$ in \eqref{e:gravitypc} to be the value that gives the best fit to our data as explained below. For comparison, the gravity model against which the radiation model is compared in Simini et al.\cite{SGMB12} also used a power law deterrence function, but had no constraints on inputs or outputs, and used nine fitted parameters (see \appref{a:gravity}{C}).

Other forms for the deterrence function in our gravity model were also investigated, but the power law in \eqref{e:gravitypc} proved the fairest comparison\cite{H19,S19}.

\subsection{The Radiation Model} \label{sec:radiation_model}

The radiation model was derived in the context of commuter flows, using the underlying assumption that a worker seeking employment will accept the most proximate job offer that meets their requirements.
The most general form of the radiation model used by Simini et al.\cite{SGMB12} is
\begin{equation}
    \Fhat_{ij} = \flowparam_i \frac{m_i n_j} {(m_i + s_{ij})(m_i + n_j + s_{ij})} \, .
    \label{e:radgeneral}
\end{equation}
The model parameter $\flowparam_i$ controls the total flow leaving each site $i$ and we have that $\sum_j \Fhat_{ij} \approx \flowparam_i$  making this radiation model a production constrained model. We will return to this approximation below.
The $n_i$ model parameter is the number of opportunities drawing commuters into site $i$, the site attractiveness parameter in this model. The $s_{ij}$ is given by the sum of all opportunities of sites closer to $i$ than $j$, the intervening opportunities measure\cite{S40}
\beq
    s_{ij} = \sum_{k | k \neq i} n_k \theta(d_{ij} - d_{ik}) \, .
\eeq
Here $\theta(x)$ is one for $x>0$ and zero otherwise so the sum does not include $n_i$ or $n_j$.
The last model parameter $m_i$ is a measure of the aspiration of commuters leaving site $i$.  That is, the larger the value of $m_i$, the greater the aspirations of the commuters leaving site $i$, and the further they must travel to achieve their aspirations. Thus, $m_i$ does not alter the total flow leaving site $i$, but $m_i$ controls the distribution of the flow leaving site $i$.

We noted above that the flow leaving each site $i$ is not exactly equal to the $t_i$ model parameter. This is easily corrected\cite{MSJB13} and by writing \eqref{e:radgeneral} using a partial fraction decomposition, we arrive at a normalised form of the radiation model
\begin{equation}
    \Fhat_{ij} = \left(\frac{N_c}{N_c-m_i}\right) \flowparam_i \frac{m_i n_j} {(m_i + s_{ij})(m_i + n_j + s_{ij})} \, .
    \label{e:radnorm}
\end{equation}
Here $N_c= \sum_i n_i$ is the total number of opportunities in the system. With this normalisation, the production constraint is perfectly enforced in the normalised radiation model, $\sum_j \Fhat_{ij} = \flowparam_i$. If $N_c \gg n_i,m_i$ then this normalised radiation model form is almost the same as \eqref{e:radgeneral} showing this correction (the factor in brackets) is often small.

One of the important features of the radiation model is that the form is fixed; there is no equivalent here to the choice of deterrence function seen in gravity models. This means there are no explicit global model parameters in the radiation model, such as the $\beta$ in \eqref{e:gravitypc}. The lack of such global model parameters (as opposed to those parameters linked to site properties) leads to the description of the radiation model as having a ``parameter-free nature''\cite{SGMB12}.

However, to use the radiation model, or indeed any spatial interaction models, we must first relate the site model parameters to values in our data. Mapping these site model parameters to data values can be done in many ways and this leads to a family of radiation models.  The versions of the radiation model analysed here are summarised in
\tabref{t:radmodels}, with more details given in \appref{A:radmodels}{B}.
In particular, the original radiation model\cite{SGMB12} used the total population $P_i$ of site $i$ to set the three site model parameters with $m_i=n_i=P_i$ and $t_i = \alpha P_i$: model F in \tabref{t:radmodels} (see also 
).
Note that $\alpha$ is a single fitted global model parameter, exemplifying how such parameters can be introduced to spatial interaction models through the mapping of data to model parameters.  In such a case, even the radiation model is no longer parameter free in the sense defined above. In our examples only our radiation models A to E are parameter free, the remaining radiation models and our gravity model both have one fitted global model parameter.

The radiation model has been widely used in the literature as the basis for a variety of other models\cite{YHEG14, liang_unraveling_2013, kang_generalized_2015}. We will focus on the family of models described above that include only minor changes to the original radiation model in order to draw conclusions about the effects of each of these changes.

\begin{table}
\caption[Different versions of the radiation model]{A summary of the different versions of the radiation model. The tick in the `Normalised?' column indicates that a model uses a normalisation that enforces the production constraint exactly  \eqref{e:radnorm}, while a cross in that column indicates that the original form \eqref{e:radgeneral} is used for that model. In each case we specify which of the site data values, ($P_i$ population, $I_i$ commuters arriving, $O_i$ commuters leaving) is used for the model site parameters (aspirations $\searchparam_i$, opportunities $\oppparam_i$, out flow $\flowparam_i$). See \appref{a:notation}{A} for a summary of the notation.
The single global model parameter $\alpha$ is found by optimising the fit of the model to the data. The model used by Simini et al.\cite{SGMB12} is equivalent to our model F. The full equations are given in \appref{A:radmodels}{B} as indicated in the final column.}
\label{t:radmodels}
\newlength{\mylength}
\settowidth{\mylength}{Normalised? }
\newlength{\mylengthtwo}
\settowidth{\mylengthtwo}{Arriving \& Departing,}
\newlength{\mylengththree}
\setlength{\mylengththree}{6pt}
\renewcommand{\arraystretch}{2.0}
\centering
\begin{tabular}{c l|c|c|c|c|c}
\multicolumn{2}{c|}{\textbf{Name}}  & $\mathbf{\searchparam_i}$ & $\mathbf{\oppparam_i}$ & $\mathbf{\flowparam_i}$ & \parbox[c]{\mylength}{\small\textbf{Normalised?}} & \textbf{Eq.}\\[\mylengththree] \hline \hline
A. & \parbox[c]{\mylengthtwo}{Total population}
                          & $\popvalue_i$     & $\popvalue_i$     & $\popvalue_i$           & $\times$
                          & (B.3) 
                          \\[\mylengththree] \hline
B. & \parbox[c]{\mylengthtwo}{Departing commuters}
                          & $\outflowvalue_i$ & $\outflowvalue_i$ & $\outflowvalue_i$       & $\times$
                          & (B.4) 
                          \\[\mylengththree] \hline
C. & \parbox[c]{\mylengthtwo}{\raggedright Departing commuters,
              Normalised} & $\outflowvalue_i$ & $\outflowvalue_i$ & $\outflowvalue_i$       & $\checkmark$
                          & (B.5) 
                          \\[\mylengththree] \hline
D. & \parbox[c]{\mylengthtwo}{\raggedright Arriving \& Departing,
         Na\"ive split}   & $\outflowvalue_i$ & $\inflowvalue_i$  & $\outflowvalue_i$       & $\times$
         & (B.6) 
         \\[\mylengththree] \hline
E. & \parbox[c]{\mylengthtwo}{\raggedright Arriving \& Departing, \newline
           Revised split} & $\inflowvalue_i$  & $\inflowvalue_i$  & $\outflowvalue_i$       & $\checkmark$
           & (B.7) 
           \\[\mylengththree] \hline
F. & \parbox[c]{\mylengthtwo}{\raggedright Total population,
        Fitted factor}    & $\popvalue_i$     & $\popvalue_i$     & $\alpha\popvalue_i$     & $\times$
        & (B.8) 
        \\[\mylengththree] \hline
G. & \parbox[c]{\mylengthtwo}{\raggedright Departing commuters,
           Fitted factor} & $\outflowvalue_i$ & $\outflowvalue_i$ & $\alpha\outflowvalue_i$ & $\times$
           & (B.9) 
           \\[\mylengththree] \hline
H. & \parbox[c]{\mylengthtwo}{\raggedright Arriving \& Departing,
     Revised, Fit factor} & $\inflowvalue_i$  & $\inflowvalue_i$  & $\alpha\outflowvalue_i$ & $\checkmark$
     & (B.10) 
     \\
\end{tabular}
\end{table}

\section{Statistical Methods} \label{s:stats}

There are two statistical challenges when dealing with spatial interaction models and data. Suitable statistical measures must be chosen to evaluate how well the models' parameters (where present) give the best fit to data, and secondly some metric must be selected to establish which model is `best'. However, the choice of this metric is not obvious. For example, one may decide to prioritise accurately predicting which pairs of sites will have zero flow ($F_{ij} = 0$) over gaining accurate estimates of the sizes of large flows. We attempt to sidestep such issues by asking in an unbiased statistical sense how \textit{probable} the models are. In order to achieve this, it is worth first considering some of the techniques found in the existing literature.

\subsection{Common techniques for comparing models \label{sec:litstats}}

A wide range of methods are used to compare spatial data against data\cite{LBR16} for a study of spatial data and models using many such measures. However, there are problems with the underlying statistical basis for many of the most popular approaches.

The S{\o}rensen-Dice coefficient is often used to compare models against real data\cite{LBR16,GLHB12,LHGD12,MSJB13,YHEG14,WOETB15,YZ19,LY20} and is sometimes referred to as the `common part of commuters' in this context. This is defined as
$\mathrm{DSC} = {\sum_{ij} \min(\Fhat_{ij}, F_{ij})}/{\sum_{ij} F_{ij}}$
for model values $\Fhat_{ij}$ and flow data $F_{ij}$ from site $i$ to site $j$. One drawback of the S{\o}rensen-Dice coefficient is that small percentage deviations in the predictions of large flows have a significant impact on the S{\o}rensen-Dice coefficient. However the main reason we do not use this measure is that it has no statistical basis; it used elsewhere because of its `intuitive explanatory power' to quote\cite{GLHB12} Gargiulo et al.\ The S{\o}rensen-Dice coefficient may still be useful but we are looking for a measure whose validity can be assessed apriori with more rigour.

Sometimes a comparison is made using statistics that assume an underlying Gaussian distribution: i.e.\ where it is assumed that the error distribution $p(F_{ij} | \Fhat_{ij})$ (the probability that the flow is found to be $F_{ij}$ given a predicted flow $\Fhat_{ij}$) is Gaussian for any $i$, $j$. A common example of a measure of this type is the coefficient of determination\cite{MSJB13,HJ16,BPTC16} $R^2= 1 - {\sum_{ij} (F_{ij} - \hat F_{ij})^2} /{\sum_{ij} (F_{ij} - \bar F)^2}$ but other examples include mean squared errors\cite{CPGB18}, and Pearson correlation coefficients\cite{LZKCJ15,GETGBMW20}. However, real data sets give integer valued data, feature no negative flows, and usually have a high proportion of very small flows. A Gaussian model of fluctuations when applied to small pairs of sites with small flows will predict real and sometimes negative flows which are poor approximations (at best) for the actual fluctuations.

The Kolmogorov-Smirnov test is also seen in spatial modelling\cite{kang_generalized_2015} and it is defined in terms of $K = \sup|\Fhat_{ij} - F_{ij}|$. One advantage is that this test does not make assumptions about the distribution of fluctuations in $F_{ij}$ or $\Fhat_{ij}$. However, the Kolmogorov-Smirnov test does require that the two input functions are independent.
Unfortunately, in spatial modelling the parameters of the model are usually estimated by fitting the model to the data so now model values $\Fhat_{ij}$ and data values $F_{ij}$ are no longer independent. The Kolmogorov-Smirnov test is then invalid and it can produce dangerous results in such circumstances\cite{steinskog_cautionary_2007}.

Finally, none of these tests measure the effects of fitting parameters: varying a model parameter to fit data can improve the accuracy of the model for that data set, but at the expense of reducing the model's predictive power on other data sets. Further discussion on these commonly used techniques, as well as an application of these techniques to the models in this paper, can be found in \appref{a:common}{D}.

\subsection{Poisson regression}

The limitations of these techniques motivate the application of alternative statistical methods\cite{ilias_bamis_constrained_2012}. Our starting point is the determination of the error distribution $p(F_{ij}| \hat F_{ij})$. Were there data on commuting for every day over a few years, we could look at the actual fluctuations in flows and examine the validity of this statistical model. However, without this data, and given that the chosen data sets (see section \ref{s:datasets}) contain discrete count data, the simplest assumption we can make is to assume that the flow $F_{ij}$ between any one pair of sites is Poisson distributed: that for any given pair of sites, we model the probability of finding flow $F_{ij}$ in the data as
$p(F_{ij} | \Fhat_{ij}) = \exp ( -\Fhat_{ij} ) (\Fhat_{ij})^{F_{ij}}/(F_{ij}!)$,
where we have taken the model estimate $\Fhat_{ij}$ to be the mean of our distribution. For small flows, the majority of values in our data, this is significantly different from a Gaussian distribution.

In fact, the models used here are built on Poisson processes making this assumption even more appropriate. We can interpret the flows given by gravity models as the flows which maximise a certain entropy function \cite{W67,W71,ES90}.  This in turn means that we can interpret a Gravity model at a microscopic level as placing discrete trips with a probability specified by the form of the entropy function. Even links with small flows are well described by a Poisson distribution in gravity models. Likewise the Radiation model \cite{SGMB12} is constructed from probabilities that commuters leaving one site will arrive at another, probabilities which are independent of the state of the system. Again the result quoted for flows in the Radiation model is just the mean of a predicted Poisson distribution.

Using these assumptions, we can now ask how probable it is that the data would be observed given the distribution predicted by the model.  This is known as `Poisson regression'.  Using Poisson regression, we calculate the log-likelihood $\ln L$ for model values $\Fhat_{ij}$, given some flow data $F_{ij}$, where we retain the option to work only with flows above a minimum value $\Fmin$, namely
\beq
\ln L(\Fmin) =
\sum_{\substack{i,j \\ i \neq j, } }
\left(
- \Fhat_{ij} + F_{ij} \ln(\Fhat_{ij}) - \ln(F_{ij}!)
\right)
\theta(F_{ij}-\Fmin)
\, .
\label{e:loglike}
\eeq
It is important here that the predicted flow in these models is never zero so we we always get a finite result for $\ln L(\Fmin)$.
Log-likelihood functions and maximum likelihood estimations provide a rigorous way to estimate fitted parameters, and to quantitatively compare how well models fit data. While adding more fitted parameters will always improve the fit of the model to the data, this risks over-fitting to the particular data set used, reducing the models' general predictive power. Thus log-likelihood values cannot tell us whether or not these fitted parameters have truly improved the model, and we need a different measure of model effectiveness.

Ideally, in order to test model effectiveness, a form of cross-validation would be used, wherein the model is fit to some data and then tested against a second data set drawn from the same distribution\cite{kohavi_study_1995}. However, the difficulty in obtaining multiple real data sets drawn from the same distribution means that some other model selection criterion must be used. One widely-used method is the Bayesian information criterion\cite{R86a, BA02a} given by
\begin{equation}
\mathrm{BIC}(\Fmin) = k \ln (n) - 2 \ln (L(\Fmin)) ,
\label{e:bic}
\end{equation}
where $k$ is the number of fitted parameters, $n$ is the number of data points, and $L$ is the likelihood. The Bayesian information criterion  can be used to compare models against a single common data set. It has a robust statistical basis\cite{S78}, introducing a penalty that increases with the number of fitted parameters. This penalty is sometimes considered too harsh\cite{V12c}.

Finally, it would be useful to have a measure of goodness-of-fit. Log-likelihood (and therefore Bayesian information criterion) values can only be used to compare models. They allow us to say one model matches real data more closely than another, but do not conclude that they resemble real data well in any absolute sense. For this, we need some value against which likelihood values can be compared. One method is to use the saturated likelihood $L_s$: the value that the likelihood would take if the predictions from the model exactly matched the data. The ratio of the actual likelihood to this saturated value must be between zero and one and can be used to define the deviance $D\geq 0$ through $L/L_s = \exp(- D/2 )$.
In our case we have that
\beq
D(\Fmin) = 2 \sum_{\substack{i,j \\ i \neq j, } }
\left(
(\Fhat_{ij} - F_{ij})
+
F_{ij}\ln( F_{ij}/\Fhat_{ij})
\right)
\theta(F_{ij}-\Fmin)
\, .
\label{e:dev}
\eeq

For all three of these statistics (log-likelihood, BIC and deviance), the lower the magnitude, the better the model fits the data.

\section{Results} \label{s:results}

\begin{figure}[htb!]
	\centering
	\includegraphics[width=\textwidth]{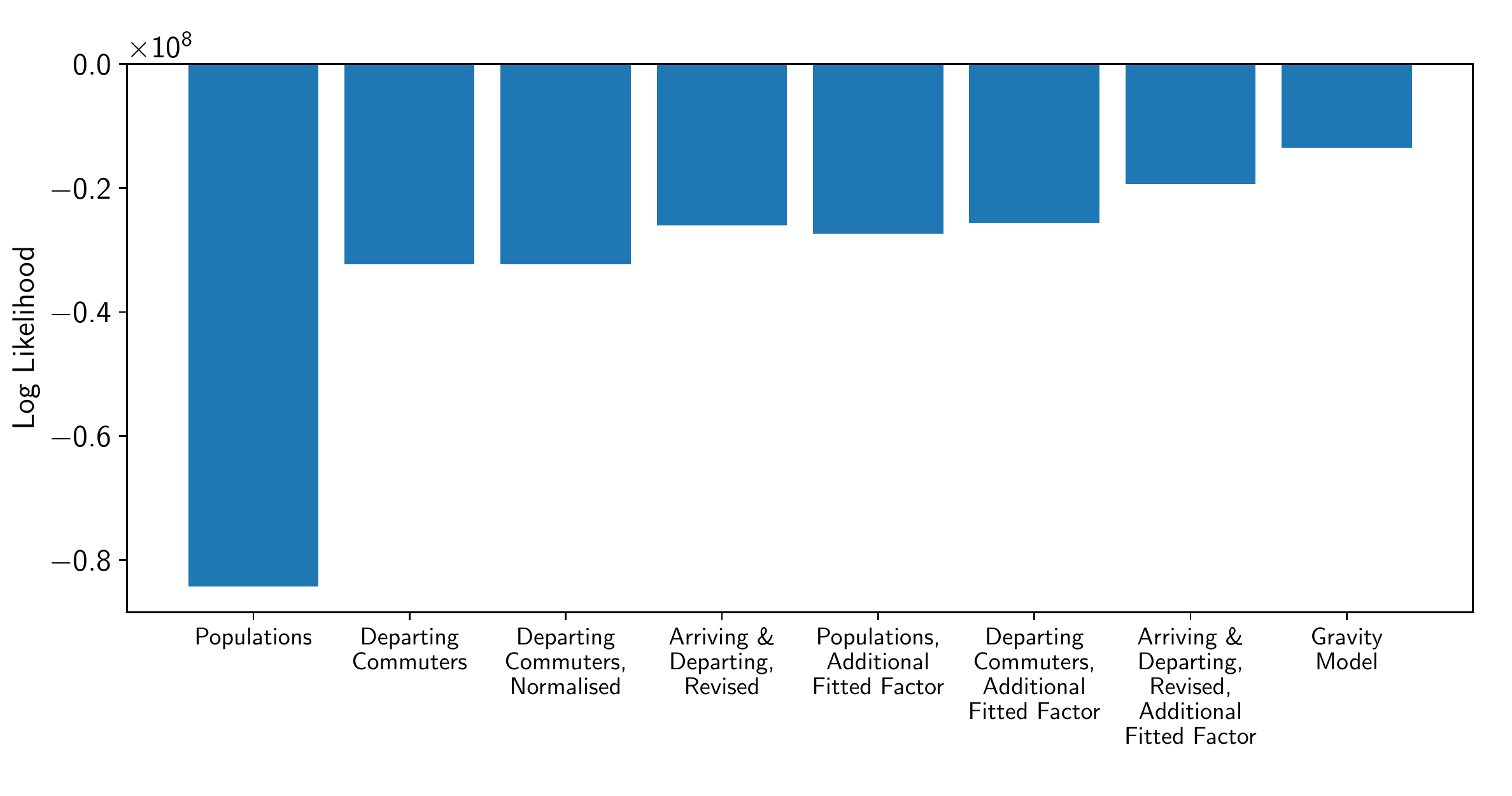} 
	\caption{The log-likelihood values \eqref{e:loglike} for radiation models A, B, C, E, F, G, H (from left to right) described in \tabref{t:radmodels}, alongside the production constrained gravity model of \eqref{e:gravitypc}. Less negative values represent better models. These data are from the US Census 2000\cite{US_census_2000}. The uncertainty in the value of any fitted parameter led to a negligible change in these results.
	}
	\label{fig:MLE_2000}
\end{figure}

Figure \ref{fig:MLE_2000} shows the log-likelihoods for the various versions of the radiation model in \tabref{t:radmodels}, and for the production constrained gravity model of \eqref{e:gravitypc}, calculated using the commuting data of the US census 2000 \cite{US_census_2000}. The exact values of the log-likelihoods and associated standard errors are shown in \tabref{table:radiation_mles}. Radiation model D (see \tabref{t:radmodels}) has been omitted from the figures in this section because of its extremely large log-likelihood --- it is far worse than any other model.
This is unsurprising since this model has assumed that $m_i$ and $n_i$ can be used analogously with $t_i$ and $n_i$ in the gravity model, without any theoretical justification for why this might be the case; an asymmetry is na\"{i}vely introduced into the model where the quantities governing site inflow and outflow are disentangled without a derivation matching this to the real world. This result thus acts as a simple check of our approach in dealing with radiation model parameters, rather than the intuitive approach of assuming that any parameters pertaining to the source site $i$ are `repulsiveness' measures and parameters pertaining to target site $j$ are `attractiveness' measures.

\begin{table}[htb!]
	\caption{The log-likelihood values \eqref{e:loglike} for the various radiation models of \tabref{t:radmodels} and the production constrained gravity model of \eqref{e:gravitypc}. The standard error in the log-likelihood comes from the uncertainty in the value of any fitted parameters, calculated from the Hessian. Thus we have no estimate of uncertainty for models without a fitted parameter, as indicated by an ``N/A'' entry. }
	\label{table:radiation_mles}
	\centering
	\begin{tabular}{|c||c|c|}\hline
		Model & Log-likelihood $\ln L$  & Error due to fit \\ \hline \hline
		A & $-8.4\times 10^7$ & N/A \\ \hline
		B & $-3.2 \times 10^7$ & N/A \\ \hline
		C & $-3.2 \times 10^7$ & N/A \\ \hline
		D & $-7.0 \times 10^8$ & N/A \\ \hline
		E & $-2.6 \times 10^7$ & N/A \\ \hline
		F & $-2.7 \times 10^7$ & $1 \times 10^{-4}$ \\ \hline
		G & $-2.6 \times 10^7$ &  $3 \times 10^{-4}$ \\ \hline
		H & $-1.9 \times 10^7$ & $3 \times 10^{-4}$ \\ \hline
		Gravity model & $-1.4 \times 10^7$ & $3 \times 10^{-4}$ \\ \hline
	\end{tabular}
\end{table}

These log-likelihoods allow for an initial comparison between models. Radiation model A (`Populations') is the worst model other than radiation model D. The total flow out of each site in radiation model A is generally significantly larger than real flows, leading to its poor performance.  Changing the site model parameters to be equal to the departing commuters data value $\outflowvalue_i$ (radiation model B - `Departing commuters') improves the model significantly, as expected. Adding in a normalisation (radiation model C - `Departing commuters, Normalised') only results in a slight improvement. This is because of the large number of commuters in the USA; the largest possible value of the normalisation factor is 1.0168 and the mean value is 1.0003.
Using a model in which the site model parameters for input and output flow, $n_i$ and $t_i$ respectively, are related to the corresponding data values, $I_i$ and $O_i$ respectively, produces the best results. This is radiation model E --- `Arriving \& Departing, Revised'.

Every model with an additional fitted factor works better than its counterpart: F is better than A, G is better than B, and H is better than E. Moreover, even model F (`Populations, Additional Fitted Factor'), which one might expect would overestimate the flows due to its large site parameter values, arrives at a better log-likelihood than either model B ('Departing Commuters') or C (`Departing Commuters - Normalised'). However, model G (`Departing Commuters, Additional Fitted Factor') is more successful than model F (`Populations, Additional Fitted Factor'), indicating that the matching of model site parameters to appropriate site data values still has merit.

The explanation for the particularly strong improvement resulting from fitting lies in the idea, corroborated below, that none of these models fit real data particularly well. Consequently, allowing a parameter to vary until the best possible value is found optimises the models' effectiveness far more than ensuring model site values are well matched to data when the overall model only approximates reality very roughly. Intriguingly, our gravity model \eqref{e:gravitypc}, whose form was chosen so as to be comparable to our radiation models,  matches our real data more closely than any of our radiation models.

\begin{figure}[htb!]
	\centering
    \includegraphics[width=\textwidth]{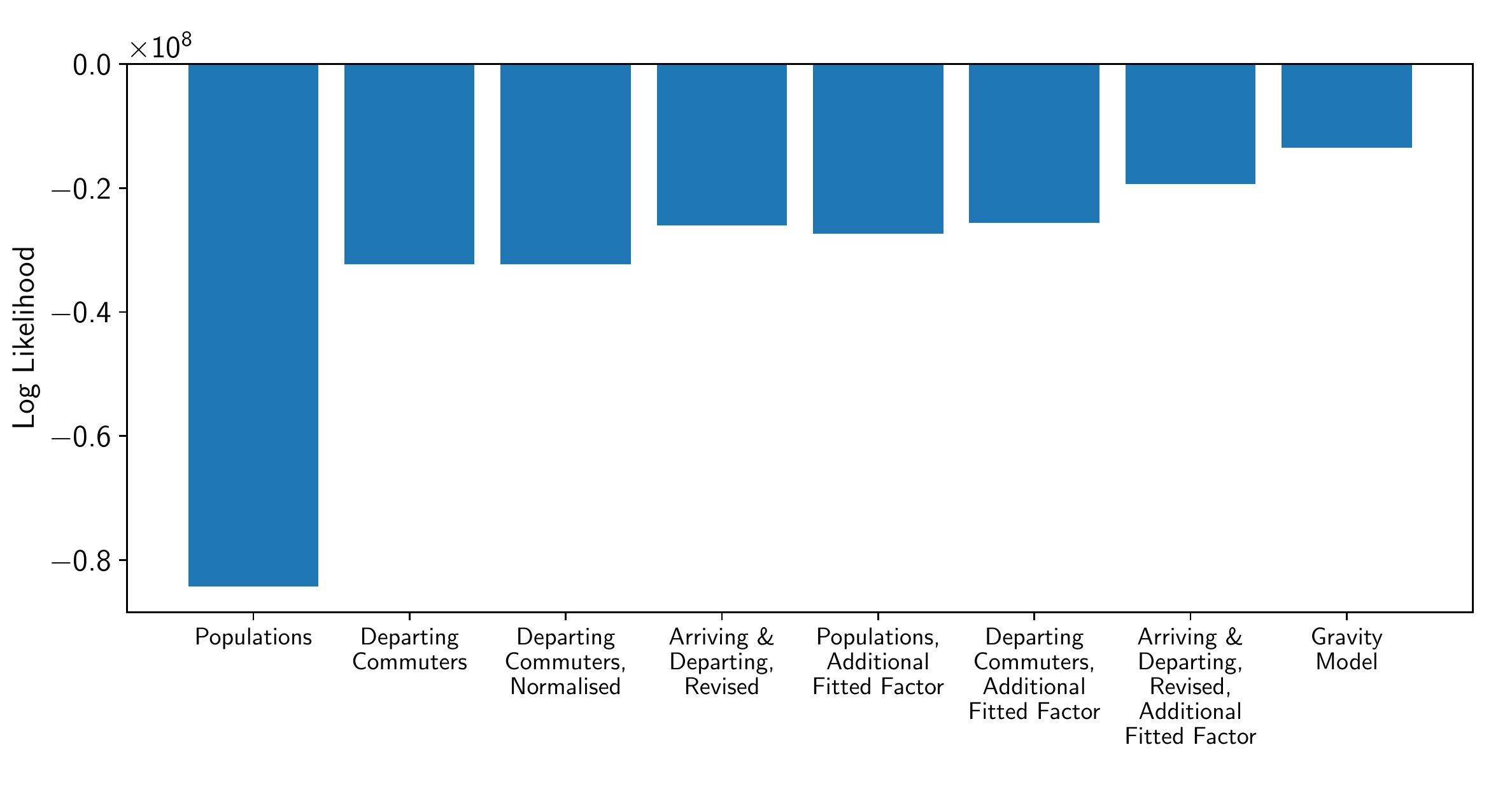} 
	\caption{Bayesian information criterion values \eqref{e:bic} for radiation models A, B, C, E, F, G, H (from left to right) described in \tabref{t:radmodels}, alongside the production constrained gravity model of \eqref{e:gravitypc}. Lower values represent better models. These data are from the US Census 2000\cite{US_census_2000}.}
	\label{fig:BIC_2000_all}
\end{figure}

Log-likelihoods alone do not tell the full story. \Figref{fig:BIC_2000_all} shows the BIC values for the models.
Despite the BIC often being regarded as overly harsh with regards to additional parameters\cite{V12c}, the trend shown is exactly the same as in Fig.\ \ref{fig:MLE_2000}. This is because the penalty applied by the BIC is $k \ln(n)$, and $\ln(n)$ is only $8.04$. This is much smaller than the log-likelihood values of order $10^7$. We can therefore conclude that there is very little risk of over-fitting, and that adding relevant additional fitted parameters significantly improves the models.

\begin{figure}[htb!]
	\centering	
    \includegraphics[width=\textwidth]{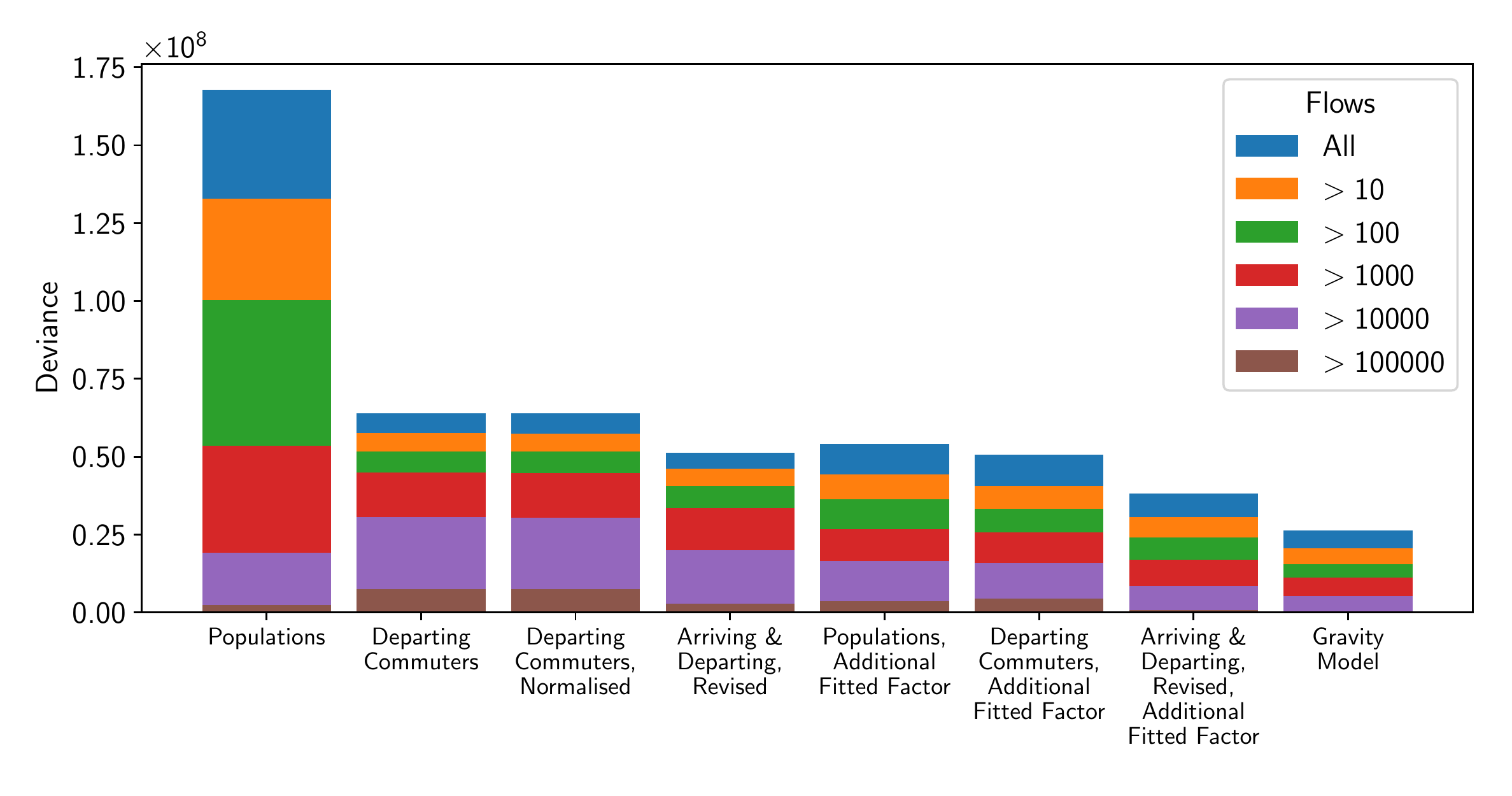} 
	\caption{Deviance values \eqref{e:dev} for radiation models A, B, C, E, F, G, H (from left to right) described in \tabref{t:radmodels}, alongside the production constrained gravity model of \eqref{e:gravitypc}, with data sets that are truncated using the minimum values shown in the legend. Lower values represent better models. These data are from the US Census 2000\cite{US_census_2000}. For each model the top of a coloured bar represents the deviance value for that model when the data is limited to flows above the value indicated in the legend.}
	\label{fig:deviance_2000}
\end{figure}

Fig.\ \ref{fig:deviance_2000} shows the deviance values for each model. The blue bars are almost identical in appearance to Fig. \ref{fig:BIC_2000_all} because the magnitude of the actual log-likelihood ($\sim 10^7$) far exceeds that of the saturated log-likelihood ($\sim 40000$). This comparison underscores how poorly these models fit real data in an absolute sense.

Given that most of the data is zero, we might wonder to what extent these trends are an artefact of how well the zero-flows are predicted rather than how well the models predict the exact sizes of the other flows. \Figref{fig:deviance_2000} addresses this by considering the deviance values for the models compared against truncated data sets, in which only flows above a certain $\Fmin$ are considered. The figure shows that the trends are almost completely as above. The only exception is for flows greater than 10,000 predicted by model A (`Populations'). This model uses the largest weights and therefore overestimates most flows, but predicts more reasonable values for the larger flows. This suggests larger flows are therefore systematically underestimated by the other models. However, only $0.022\%$ of flows predicted by model A are greater than 10,000, so this trend does not significantly affect the validity of our overall conclusions.

\begin{figure}[htb!]
	\centering	
    \includegraphics[width=\textwidth]{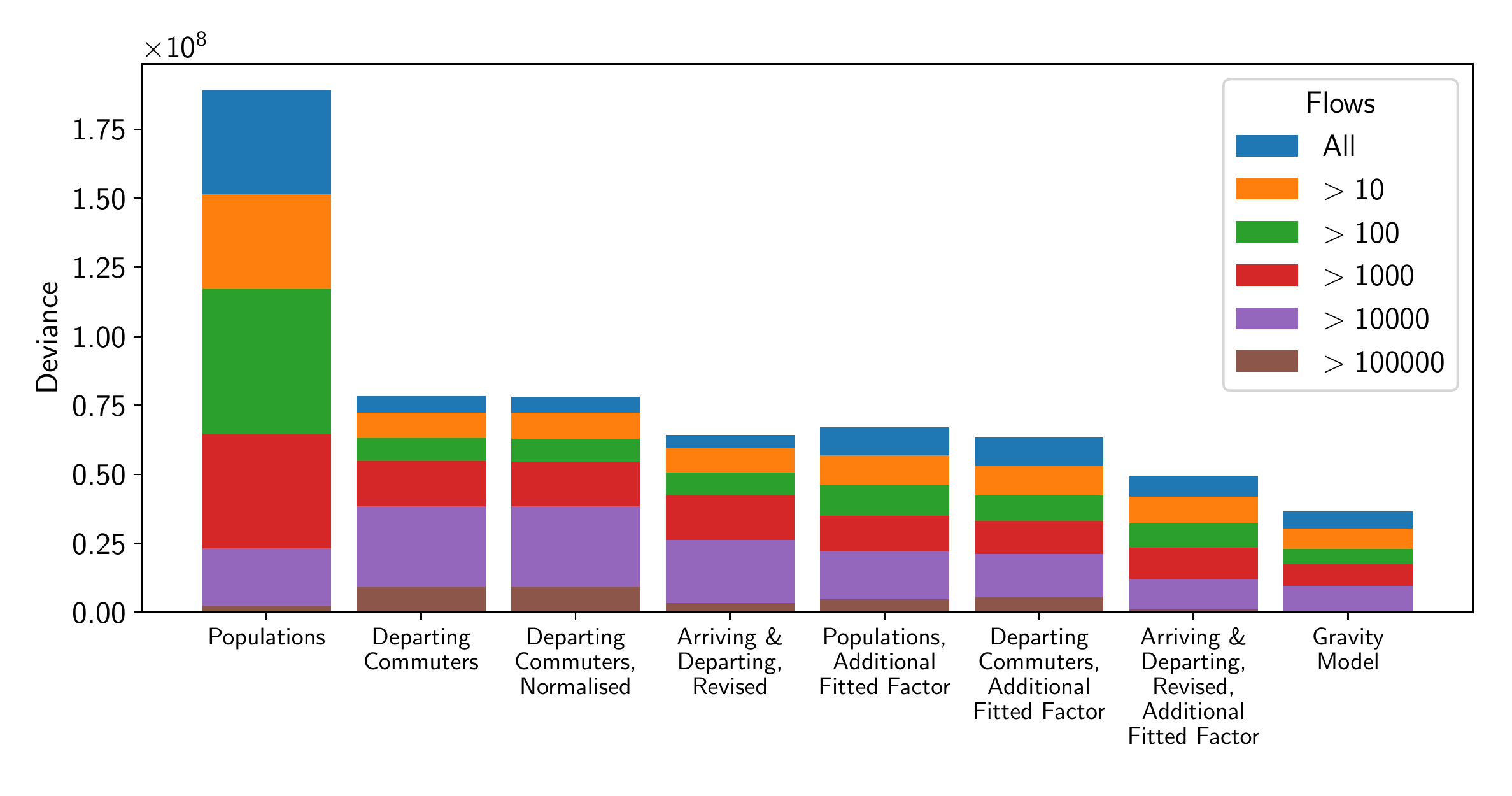}  
	\caption{Deviance values \eqref{e:dev} for radiation models A, B, C, E, F, G, H (from left to right) described in \tabref{t:radmodels}, alongside the production constrained gravity model of \eqref{e:gravitypc},  with data sets that are truncated using the minimum values shown in the legend. Lower values represent better models. These data are from the American Commuter Survey 2009-2013 \cite{ACS_2013}. For each model the top of a coloured bar represents the deviance value for that model when the data is limited to flows above the value indicated in the legend.}
	\label{fig:deviance_acs}
\end{figure}

Finally, we consider the second data set (the American Commuter Survey\cite{ACS_2013} 2009-2013). In \figref{fig:deviance_acs} we show the deviance values for these data, though the trends are the same in all our measures. The results for this data reinforce all of our conclusions.

\section{Conclusions and Discussion}

For our data on modern US commuter flows, the most accurate flow predictions came from the production-constrained gravity model. Looking at the truly ``parameter free'' radiation models, that is models with no fitted global parameters, radiation model E (`Arriving \& Departing, Revised') was most successful. The set of parameter free radiation models A--E showed that matching each model parameter to an appropriate data value improves the model performance as we should expect.
This radiation model E benefits from a number of improvements over the original radiation model: choosing an appropriate input data set (e.g.\ the number of individuals who leave each site rather than population); correctly adjusting the model to include measures of both attractiveness and repulsiveness for each site; and introducing the correct normalisation.

Another conclusion was that adding an additional global parameter, and setting that parameter by finding the best fit, improves the performance of any model.  The penalty of having an extra parameter is negligible for our data sets while there is vast room for improvement in what are poor fits in statistical terms. This is why the radiation model that best fits both data sets is radiation model H. This is the same as radiation model E, but with a single additional fitted parameter.

Despite these improvements, and in direct contrast with the results elsewhere\cite{SGMB12}, our statistical measures show that for these US commuting data sets the radiation model is vastly inferior to an appropriately chosen gravity model for most realistic purposes, i.e.\ where there is data that can be used to fit parameters -- what appears to be a small visual difference between models in our plots represents a large numerical difference.

The relative success of our chosen gravity model highlights another result. The use of a gravity model on the same data\cite{SGMB12} made less successful predictions than the radiation model in spite of its having nine fitted parameters to the latter's zero. This underscores the importance of constraints, and the requirement that only models with corresponding constraints be compared against each other when the impact of these constraints is not the topic of investigation. This is why in this work all our models are production constrained in order to make our comparisons fair.

By examining the deviance values, we further established that none of these models fit our data well in an absolute sense. This is unsurprising: the large number of factors affecting commuter flows --- geographical and socio-economic --- limit the extent to which a simple model with very few parameters could make accurate predictions.

Our work leads us to make recommendations for spatial interaction modelling in general.
First, we suggest that non-Gaussian regression (in particular Poisson regression) as applied to log-likelihood, Bayesian information criterion and deviance, are good statistical methods to use when analysing spatial interaction models. These have a firm theoretical grounding and provide an unbiased statistical approach.
Second, we should ensure any feature that is not being explicitly tested is controlled for. Here, this means all our models enforce the production constraint. In fact, it would be trivial to add the input constraint into all these models, as is standard for gravity models\cite{ES90}. Such an improvement requires no additional parameters.
Third, the small penalty in the Bayesian information criterion arising from additional parameters, as well as the lower deviance values of models with fitted parameters, attest to the fact that if data exist that can be used for fitting, then a model with many physically relevant parameters can be improved by fitting to this data. Having such fitted model parameters is an advantage, not a disadvantage.
Fourth, models should make use of as much available information as possible. We found that if we used the actual commuter flows in and out of sites in a way that matched that narrative behind a model, then results were better than trying to use the population as some proxy for the actual flows.
Lastly, these simple spatial interaction models should be used only to provide an outline of real-world processes, with fitted parameter values giving general insights into spatially-constrained processes. These models are only ever crude approximations of reality.

\section*{Acknowledgements}

TSE thanks E.\ Bamis, P.\ Expert, M.\ T.\ Gastner, and R.\ J.\ Rivers for useful conversations.

\section*{Author contributions statement}

BH and APS are equal first authors. BH and APS conducted the numerical simulations and data analysis. BH, APS and TSE analysed the results, interpreted the results, and wrote the manuscript.

\section*{Additional information}

The authors declare no competing interests.
All the data used in this work is publicly available as cited within the text\cite{US_census_2000,ACS_2013,distance_data,cest_2000,cest_1990}.

%
%




\begin{thebibliography}{10}
\urlstyle{rm}
\expandafter\ifx\csname url\endcsname\relax
  \def\url#1{\texttt{#1}}\fi
\expandafter\ifx\csname urlprefix\endcsname\relax\def\urlprefix{URL }\fi
\expandafter\ifx\csname doiprefix\endcsname\relax\def\doiprefix{DOI: }\fi
\providecommand{\bibinfo}[2]{#2}
\providecommand{\eprint}[2][]{\url{#2}}
\providecommand{\JournalTitle}[1]{\emph{#1}}

\bibitem{US_census_2000}
\bibinfo{author}{{\relax United States Census Bureau}}.
\newblock \bibinfo{title}{County-to-county worker flow files}
  (\bibinfo{year}{2001}).

\bibitem{SGMB12}
\bibinfo{author}{Simini, F.}, \bibinfo{author}{Gonzalez, M.~C.},
  \bibinfo{author}{Maritan, A.} \& \bibinfo{author}{Barabasi, A.-L.}
\newblock \bibinfo{journal}{\bibinfo{title}{A universal model for mobility and
  migration patterns}}.
\newblock {\emph{\JournalTitle{Nature}}} \textbf{\bibinfo{volume}{484}},
  \bibinfo{pages}{96--100}, \doiprefix\url{10.1038/nature10856}
  (\bibinfo{year}{2012}).

\bibitem{ACS_2013}
\bibinfo{author}{{\relax United States Census Bureau}}.
\newblock \bibinfo{title}{2009-2013 {5-Year American Community Survey Commuting
  Flows}} (\bibinfo{year}{2013}).

\bibitem{cest_1990}
\bibinfo{author}{{\relax United States Census Bureau}}.
\newblock \bibinfo{title}{State and county intercensal tables: 1990-2000}
  (\bibinfo{year}{2016}).

\bibitem{cest_2000}
\bibinfo{author}{{\relax United States Census Bureau}}.
\newblock \bibinfo{title}{County intercensal tables: 2000-2010}
  (\bibinfo{year}{2017}).

\bibitem{distance_data}
\bibinfo{author}{{\relax National Bureau of Economic Research}}.
\newblock \bibinfo{title}{County distance database} (\bibinfo{year}{2016}).

\bibitem{ES90}
\bibinfo{author}{Erlander, S.} \& \bibinfo{author}{Stewart, N.}
\newblock \emph{\bibinfo{title}{The Gravity Model in Transportation Analysis}}
  (\bibinfo{publisher}{VSP}, \bibinfo{year}{1990}).

\bibitem{NR12}
\bibinfo{author}{Nijkamp, P.} \& \bibinfo{author}{Reggiani, A.}
\newblock \emph{\bibinfo{title}{Interaction, evolution and chaos in space}}
  (\bibinfo{publisher}{Springer Science \& Business Media},
  \bibinfo{year}{2012}).

\bibitem{balcan_multiscale_2009}
\bibinfo{author}{Balcan, D.} \emph{et~al.}
\newblock \bibinfo{journal}{\bibinfo{title}{Multiscale mobility networks and
  the spatial spreading of infectious diseases}}.
\newblock {\emph{\JournalTitle{Proceedings of the National Academy of
  Sciences}}} \textbf{\bibinfo{volume}{106}}, \bibinfo{pages}{21484--21489}
  (\bibinfo{year}{2009}).

\bibitem{kaluza_complex_2010}
\bibinfo{author}{Kaluza, P.}, \bibinfo{author}{K\"{o}lzsch, A.},
  \bibinfo{author}{Gastner, M.~T.} \& \bibinfo{author}{Blasius, B.}
\newblock \bibinfo{journal}{\bibinfo{title}{The complex network of global cargo
  ship movements}}.
\newblock {\emph{\JournalTitle{Journal of The Royal Society Interface}}}
  \textbf{\bibinfo{volume}{7}}, \bibinfo{pages}{1093--1103}
  (\bibinfo{year}{2010}).

\bibitem{viboud_synchrony_2006}
\bibinfo{author}{Viboud, C.} \emph{et~al.}
\newblock \bibinfo{journal}{\bibinfo{title}{Synchrony, {Waves}, and {Spatial}
  {Hierarchies} in the {Spread} of {Influenza}}}.
\newblock {\emph{\JournalTitle{Science}}} \textbf{\bibinfo{volume}{312}},
  \bibinfo{pages}{447--451} (\bibinfo{year}{2006}).

\bibitem{W67}
\bibinfo{author}{Wilson, A.~G.}
\newblock \bibinfo{journal}{\bibinfo{title}{A statistical theory of spatial
  distribution models}}.
\newblock {\emph{\JournalTitle{Transportation Research}}}
  \textbf{\bibinfo{volume}{1}}, \bibinfo{pages}{253--269}
  (\bibinfo{year}{1967}).

\bibitem{W71}
\bibinfo{author}{Wilson, A.~G.}
\newblock \bibinfo{journal}{\bibinfo{title}{A {Family} of {Spatial}
  {Interaction} {Models}, and {Associated} {Developments}}}.
\newblock {\emph{\JournalTitle{Environment and Planning A: Economy and Space}}}
  \textbf{\bibinfo{volume}{3}}, \bibinfo{pages}{1--32} (\bibinfo{year}{1971}).

\bibitem{H19}
\bibinfo{author}{Hilton, B.}
\newblock \emph{\bibinfo{title}{Investigations Into the Accuracy of Spatial
  Interaction Models}}.
\newblock Master's thesis, \bibinfo{school}{Imperial College London},
  \doiprefix\url{10.6084/m9.figshare.9752504} (\bibinfo{year}{2019}).

\bibitem{S19}
\bibinfo{author}{Sood, A.~P.}
\newblock \emph{\bibinfo{title}{An Investigation of Models of Flow in Complete
  Spatially Embedded Networks}}.
\newblock Master's thesis, \bibinfo{school}{Imperial College London},
  \doiprefix\url{10.6084/m9.figshare.9751919} (\bibinfo{year}{2019}).

\bibitem{S40}
\bibinfo{author}{Stouffer, S.~A.}
\newblock \bibinfo{journal}{\bibinfo{title}{Intervening opportunities: A theory
  relating to mobility and distance}}.
\newblock {\emph{\JournalTitle{American Sociological Review}}}
  \textbf{\bibinfo{volume}{5}}, \bibinfo{pages}{845--867},
  \doiprefix\url{10.2307/2084520.} (\bibinfo{year}{1940}).

\bibitem{MSJB13}
\bibinfo{author}{Masucci, A.~P.}, \bibinfo{author}{Serras, J.},
  \bibinfo{author}{Johansson, A.} \& \bibinfo{author}{Batty, M.}
\newblock \bibinfo{journal}{\bibinfo{title}{Gravity versus radiation models: On
  the importance of scale and heterogeneity in commuting flows}}.
\newblock {\emph{\JournalTitle{Physical Review E}}}
  \textbf{\bibinfo{volume}{88}}, \bibinfo{pages}{022812},
  \doiprefix\url{10.1103/PhysRevE.88.022812} (\bibinfo{year}{2013}).

\bibitem{YHEG14}
\bibinfo{author}{Yang, Y.}, \bibinfo{author}{Herrera, C.},
  \bibinfo{author}{Eagle, N.} \& \bibinfo{author}{Gonz{\'{a}}lez, M.~C.}
\newblock \bibinfo{journal}{\bibinfo{title}{Limits of predictability in
  commuting flows in the absence of data for calibration}}.
\newblock {\emph{\JournalTitle{Scientific Reports}}}
  \textbf{\bibinfo{volume}{4}}, \bibinfo{pages}{5662},
  \doiprefix\url{10.1038/srep05662} (\bibinfo{year}{2014}).

\bibitem{liang_unraveling_2013}
\bibinfo{author}{Liang, X.}, \bibinfo{author}{Zhao, J.}, \bibinfo{author}{Dong,
  L.} \& \bibinfo{author}{Xu, K.}
\newblock \bibinfo{journal}{\bibinfo{title}{Unraveling the origin of
  exponential law in intra-urban human mobility}}.
\newblock {\emph{\JournalTitle{Scientific Reports}}}
  \textbf{\bibinfo{volume}{3}}, \bibinfo{pages}{2983} (\bibinfo{year}{2013}).

\bibitem{kang_generalized_2015}
\bibinfo{author}{Kang, C.}, \bibinfo{author}{Liu, Y.}, \bibinfo{author}{Guo,
  D.} \& \bibinfo{author}{Qin, K.}
\newblock \bibinfo{journal}{\bibinfo{title}{A {Generalized} {Radiation} {Model}
  for {Human} {Mobility}: {Spatial} {Scale}, {Searching} {Direction} and {Trip}
  {Constraint}}}.
\newblock {\emph{\JournalTitle{PLOS One}}} \textbf{\bibinfo{volume}{10}}
  (\bibinfo{year}{2015}).

\bibitem{LBR16}
\bibinfo{author}{Lenormand, M.}, \bibinfo{author}{Bassolas, A.} \&
  \bibinfo{author}{Ramasco, J.~J.}
\newblock \bibinfo{journal}{\bibinfo{title}{Systematic comparison of trip
  distribution laws and models}}.
\newblock {\emph{\JournalTitle{Journal of Transport Geography}}}
  \textbf{\bibinfo{volume}{51}}, \bibinfo{pages}{158--169},
  \doiprefix\url{10.1016/j.jtrangeo.2015.12.008} (\bibinfo{year}{2016}).

\bibitem{GLHB12}
\bibinfo{author}{Gargiulo, F.}, \bibinfo{author}{Lenormand, M.},
  \bibinfo{author}{Huet, S.} \& \bibinfo{author}{Baqueiro~Espinosa, O.}
\newblock \bibinfo{journal}{\bibinfo{title}{Commuting network models: Getting
  the essentials}}.
\newblock {\emph{\JournalTitle{Journal of Artificial Societies and Social
  Simulation}}} \textbf{\bibinfo{volume}{15}}, \bibinfo{pages}{6--}
  (\bibinfo{year}{2012}).

\bibitem{LHGD12}
\bibinfo{author}{Lenormand, M.}, \bibinfo{author}{Huet, S.},
  \bibinfo{author}{Gargiulo, F.} \& \bibinfo{author}{Deffuant, G.}
\newblock \bibinfo{journal}{\bibinfo{title}{A universal model of commuting
  networks}}.
\newblock {\emph{\JournalTitle{PLOS ONE}}} \textbf{\bibinfo{volume}{7}},
  \bibinfo{pages}{e45985}, \doiprefix\url{10.1371/journal.pone.0045985}
  (\bibinfo{year}{2012}).
\newblock \eprint{1203.5184v2}.

\bibitem{WOETB15}
\bibinfo{author}{Wesolowski, A.}, \bibinfo{author}{O'Meara, W.~P.},
  \bibinfo{author}{Eagle, N.}, \bibinfo{author}{Tatem, A.~J.} \&
  \bibinfo{author}{Buckee, C.~O.}
\newblock \bibinfo{journal}{\bibinfo{title}{Evaluating spatial interaction
  models for regional mobility in sub-saharan africa}}.
\newblock {\emph{\JournalTitle{{PLOS} Computational Biology}}}
  \textbf{\bibinfo{volume}{11}}, \bibinfo{pages}{e1004267},
  \doiprefix\url{10.1371/journal.pcbi.1004267} (\bibinfo{year}{2015}).

\bibitem{KLGQ15}
\bibinfo{author}{Kang, C.}, \bibinfo{author}{Liu, Y.}, \bibinfo{author}{Guo,
  D.} \& \bibinfo{author}{Qin, K.}
\newblock \bibinfo{journal}{\bibinfo{title}{A generalized radiation model for
  human mobility: Spatial scale, searching direction and trip constraint}}.
\newblock {\emph{\JournalTitle{{PLOS} {ONE}}}} \textbf{\bibinfo{volume}{10}},
  \bibinfo{pages}{e0143500}, \doiprefix\url{10.1371/journal.pone.0143500}
  (\bibinfo{year}{2015}).

\bibitem{GSSHSVSBR17}
\bibinfo{author}{Grauwin, S.} \emph{et~al.}
\newblock \bibinfo{journal}{\bibinfo{title}{Identifying and modeling the
  structural discontinuities of human interactions}}.
\newblock {\emph{\JournalTitle{Scientific Reports}}}
  \textbf{\bibinfo{volume}{7}}, \doiprefix\url{10.1038/srep46677}
  (\bibinfo{year}{2017}).

\bibitem{YGZMWL20}
\bibinfo{author}{Yao, X.} \emph{et~al.}
\newblock \bibinfo{journal}{\bibinfo{title}{Origin-destination flow, data
  imputation, spatial interaction network, graph embedding, graph convolution;
  spatial origin-destination flow imputation using graph convolutional
  networks}}.
\newblock {\emph{\JournalTitle{{IEEE} Transactions on Intelligent
  Transportation Systems}}} \bibinfo{pages}{1--11},
  \doiprefix\url{10.1109/tits.2020.3003310} (\bibinfo{year}{2020}).



\bibitem{YZ19}
\bibinfo{author}{Yan, X.-Y.} \& \bibinfo{author}{Zhou, T.}
\newblock \bibinfo{journal}{\bibinfo{title}{Destination choice game: A spatial
  interaction theory on human mobility}}.
\newblock {\emph{\JournalTitle{Scientific Reports}}}
  \textbf{\bibinfo{volume}{9}}, \doiprefix\url{10.1038/s41598-019-46026-w}
  (\bibinfo{year}{2019}).

\bibitem{LY20}
\bibinfo{author}{Liu, E.-J.} \& \bibinfo{author}{Yan, X.-Y.}
\newblock \bibinfo{journal}{\bibinfo{title}{A universal opportunity model for
  human mobility}}.
\newblock {\emph{\JournalTitle{Scientific Reports}}}
  \textbf{\bibinfo{volume}{10}}, \bibinfo{pages}{4657},
  \doiprefix\url{10.1038/s41598-020-61613-y} (\bibinfo{year}{2020}).

\bibitem{HJ16}
\bibinfo{author}{Hong, I.} \& \bibinfo{author}{Jung, W.-S.}
\newblock \bibinfo{journal}{\bibinfo{title}{Application of gravity model on the
  korean urban bus network}}.
\newblock {\emph{\JournalTitle{Physica A: Statistical Mechanics and its
  Applications}}} \textbf{\bibinfo{volume}{462}}, \bibinfo{pages}{48--55},
  \doiprefix\url{10.1016/j.physa.2016.06.055} (\bibinfo{year}{2016}).

\bibitem{BPTC16}
\bibinfo{author}{Beir{\'{o}}, M.~G.}, \bibinfo{author}{Panisson, A.},
  \bibinfo{author}{Tizzoni, M.} \& \bibinfo{author}{Cattuto, C.}
\newblock \bibinfo{journal}{\bibinfo{title}{Predicting human mobility through
  the assimilation of social media traces into mobility models}}.
\newblock {\emph{\JournalTitle{{EPJ} Data Science}}}
  \textbf{\bibinfo{volume}{5}}, \doiprefix\url{10.1140/epjds/s13688-016-0092-2}
  (\bibinfo{year}{2016}).

\bibitem{CPGB18}
\bibinfo{author}{Curiel, R.~P.}, \bibinfo{author}{Pappalardo, L.},
  \bibinfo{author}{Gabrielli, L.} \& \bibinfo{author}{Bishop, S.~R.}
\newblock \bibinfo{journal}{\bibinfo{title}{Gravity and scaling laws of city to
  city migration}}.
\newblock {\emph{\JournalTitle{{PLOS} {ONE}}}} \textbf{\bibinfo{volume}{13}},
  \bibinfo{pages}{e0199892}, \doiprefix\url{10.1371/journal.pone.0199892}
  (\bibinfo{year}{2018}).

\bibitem{LZKCJ15}
\bibinfo{author}{Liu, J.}, \bibinfo{author}{Zhao, K.}, \bibinfo{author}{Khan,
  S.}, \bibinfo{author}{Cameron, M.} \& \bibinfo{author}{Jurdak, R.}
\newblock \bibinfo{title}{Multi-scale population and mobility estimation with
  geo-tagged tweets}.
\newblock In \emph{\bibinfo{booktitle}{2015 31st {IEEE} International
  Conference on Data Engineering Workshops}},
  \doiprefix\url{10.1109/icdew.2015.7129551} (\bibinfo{publisher}{{IEEE}},
  \bibinfo{year}{2015}).

\bibitem{GETGBMW20}
\bibinfo{author}{Giles, J.~R.} \emph{et~al.}
\newblock \bibinfo{journal}{\bibinfo{title}{The duration of travel impacts the
  spatial dynamics of infectious diseases}}.
\newblock {\emph{\JournalTitle{Proceedings of the National Academy of
  Sciences}}} \bibinfo{pages}{201922663},
  \doiprefix\url{10.1073/pnas.1922663117} (\bibinfo{year}{2020}).

\bibitem{steinskog_cautionary_2007}
\bibinfo{author}{Steinskog, D.~J.}, \bibinfo{author}{Tj{\o}stheim, D.~B.} \&
  \bibinfo{author}{Kvamst{\o}, N.~G.}
\newblock \bibinfo{journal}{\bibinfo{title}{A cautionary note on the use of the
  {Kolmogorov}--{Smirnov} test for normality}}.
\newblock {\emph{\JournalTitle{Monthly Weather Review}}}
  \textbf{\bibinfo{volume}{135}}, \bibinfo{pages}{1151--1157}
  (\bibinfo{year}{2007}).

\bibitem{ilias_bamis_constrained_2012}
\bibinfo{author}{{I. Bamis}}.
\newblock \emph{\bibinfo{title}{Constrained {Gravity} {Models} for {Network}
  {Flows}}}.
\newblock \bibinfo{type}{{MSc} {Thesis}}, \bibinfo{school}{Imperial College},
  \bibinfo{address}{London} (\bibinfo{year}{2012}).

\bibitem{kohavi_study_1995}
\bibinfo{author}{Kohavi, R.}
\newblock \bibinfo{title}{A study of cross-validation and bootstrap for
  accuracy estimation and model selection}.
\newblock In \emph{\bibinfo{booktitle}{{International} {Joint} {Conference} on
  {Artificial} {Intelligence} (IJCAI)}}, vol.~\bibinfo{volume}{14},
  \bibinfo{pages}{1137--1145} (\bibinfo{organization}{Montreal, Canada},
  \bibinfo{year}{1995}).

\bibitem{R86a}
\bibinfo{author}{Raftery, A.~E.}
\newblock \bibinfo{journal}{\bibinfo{title}{Choosing models for
  cross-classifications}}.
\newblock {\emph{\JournalTitle{{American Sociological Review}}}}
  \textbf{\bibinfo{volume}{51}}, \bibinfo{pages}{145--146}
  (\bibinfo{year}{1986}).

\bibitem{BA02a}
\bibinfo{author}{Burnham, K.~P.} \& \bibinfo{author}{Anderson, D.~R.}
\newblock \emph{\bibinfo{title}{Model Selection and Multimodel Inference: A
  Practical Information-Theoretic Approach}}
  (\bibinfo{publisher}{Springer-Verlag New York}, \bibinfo{year}{2002}).

\bibitem{S78}
\bibinfo{author}{Schwarz, G.}
\newblock \bibinfo{journal}{\bibinfo{title}{Estimating the dimension of a
  model}}.
\newblock {\emph{\JournalTitle{The Annals of Statistics}}}
  \textbf{\bibinfo{volume}{6}}, \bibinfo{pages}{461--464}
  (\bibinfo{year}{1978}).

\bibitem{V12c}
\bibinfo{author}{Vrieze, S.~I.}
\newblock \bibinfo{journal}{\bibinfo{title}{Model selection and psychological
  theory: A discussion of the differences between the {Akaike} information
  criterion ({AIC}) and the {Bayesian} information criterion ({BIC})}}.
\newblock {\emph{\JournalTitle{Psychological Methods}}}
  \textbf{\bibinfo{volume}{17}}, \bibinfo{pages}{228--243},
  \doiprefix\url{10.1037/a0027127} (\bibinfo{year}{2012}).

\end{thebibliography}


%
%


\renewcommand{\thesection}{\Alph{section}}
\renewcommand{\thefigure}{\thesection\arabic{figure}}
\renewcommand{\thetable}{\thesection\arabic{table}}
\setcounter{section}{0}
\setcounter{figure}{0}
\setcounter{table}{0}
\renewcommand{\theHsection}{\Alph{section}}
\renewcommand{\theHequation}{\thesection\arabic{equation}}
\renewcommand{\theHfigure}{\thesection\arabic{figure}}
\renewcommand{\theHtable}{\thesection\arabic{table}}

%


\section{Summary of Notation}\label{a:notation}

A summary of the notation used in this work is given in \tabref{t:param}.
\begin{table}[htb]
\centering
\begin{tabular}{c|p{0.8\textwidth}}
  Notation & Meaning \\ \hline\hline
  $i$, $j$          & Indices of sites.\\
  $\popvalue_i$     & The population of site $i$. \\
  $\outflowvalue_i$ & The number of commuters leaving site $i$. \\
  $\inflowvalue_i$  & The number of commuters arriving at site $i$. \\
  $N_c$             & The total number of commuters in the data.
                      This satisfies
                      $N_c  = \sum_i \outflowvalue_i = \sum_j \inflowvalue_j$. \\
  $F_{ij}$          & The actual flow from a source site $i$ to a target site $j$ as found in the data. \\
  $d_{ij}$          & A measure of the distance from site $i$ to site $j$.
  \\[6pt] \hline
  \rule{0pt}{15pt} $\Fhat_{ij}$      & The estimated flow from a source site $i$ to a target site $j$ as predicted by some model. \\
  $w_i$				& The site `weight' model parameter. Controls the flow into and out of a site. \\
  $m_i$    		    & The site `aspiration' model parameter. Controls the distribution of flows from site $i$. \\
  $n_i$  			& The site `attractiveness' model parameter, the number of `opportunities' . Controls flow into site $i$.\\
  $\flowparam_i$    & Model parameter controlling the total flow leaving site $i$ (site `repulsiveness'). \\[6pt]
\end{tabular}

\caption[Parameter Summary]{A summary of the different data values (top half) and the different model parameters (bottom half) used in this paper.}
\label{t:param}
\end{table}

\clearpage
\section{Versions of the Radiation model}\label{A:radmodels}

In this section we give explicit forms for the Radiation models used in our work  written in terms of the actual data values used. These are summarised in \tabref{t:radmodelsapp} (reproduced from the main text) with detailed equation given in the following subsections.
In each case we explain how the parameters $m_i$, $n_j$ and $t_i$ of the Radiation model are replaced by values obtained with data.  For simplicity, we will repeat here our standard form for the Radiation model:-
\begin{equation}
    \Fhat_{ij} = \flowparam_i \frac{m_i n_j} {(m_i + s_{ij})(m_i + n_j + s_{ij})} \, .
    \label{e:radgeneralapp}
\end{equation}
We will also note any normalisation factors used \cite{MSJB13}, something not included in the simple form \eqref{e:radgeneralapp} above. For completeness this is
\begin{equation}
    \Fhat_{ij} = \left(\frac{N_c}{N_c-m_i}\right) \flowparam_i \frac{m_i n_j} {(m_i + s_{ij})(m_i + n_j + s_{ij})} \, .
    \label{e:radnormapp}
\end{equation}
Here $N_c= \sum_i n_i$ is the total number of opportunities in the system.

\begin{table}[htb]
\settowidth{\mylength}{Normalised? }
\settowidth{\mylengthtwo}{Arriving \& Departing,}
\setlength{\mylengththree}{6pt}
\renewcommand{\arraystretch}{2.0}
\centering
\begin{tabular}{c l|c|c|c|c|c}
\multicolumn{2}{c|}{\textbf{Name}}  & $\mathbf{\searchparam_i}$ & $\mathbf{\oppparam_i}$ & $\mathbf{\flowparam_i}$ & \parbox[c]{\mylength}{\small\textbf{Normalised?}} & \textbf{Eq.}\\[\mylengththree] \hline \hline
A. & \parbox[c]{\mylengthtwo}{Total population}
                          & $\popvalue_i$     & $\popvalue_i$     & $\popvalue_i$           & $\times$
                          & \eqref{A:populations_modelA}
                          \\[\mylengththree] \hline
B. & \parbox[c]{\mylengthtwo}{Departing commuters}
                          & $\outflowvalue_i$ & $\outflowvalue_i$ & $\outflowvalue_i$       & $\times$
                          & \eqref{A:departing_modelB}
                          \\[\mylengththree] \hline
C. & \parbox[c]{\mylengthtwo}{\raggedright Departing commuters,
              Normalised} & $\outflowvalue_i$ & $\outflowvalue_i$ & $\outflowvalue_i$       & $\checkmark$
                          & \eqref{A:departing_normed_modelC}
                          \\[\mylengththree] \hline
D. & \parbox[c]{\mylengthtwo}{\raggedright Arriving \& Departing,
         Na\"ive split}   & $\outflowvalue_i$ & $\inflowvalue_i$  & $\outflowvalue_i$       & $\times$
         & \eqref{A:naive_modelD}
         \\[\mylengththree] \hline
E. & \parbox[c]{\mylengthtwo}{\raggedright Arriving \& Departing, \newline
           Revised split} & $\inflowvalue_i$  & $\inflowvalue_i$  & $\outflowvalue_i$       & $\checkmark$
           & \eqref{A:revised_modelE}
           \\[\mylengththree] \hline
F. & \parbox[c]{\mylengthtwo}{\raggedright Total population,
        Fitted factor}    & $\popvalue_i$     & $\popvalue_i$     & $\alpha\popvalue_i$     & $\times$
        & \eqref{A:departing_modelF}
        \\[\mylengththree] \hline
G. & \parbox[c]{\mylengthtwo}{\raggedright Departing commuters,
           Fitted factor} & $\outflowvalue_i$ & $\outflowvalue_i$ & $\alpha\outflowvalue_i$ & $\times$
           & \eqref{A:departing_modelG}
           \\[\mylengththree] \hline
H. & \parbox[c]{\mylengthtwo}{\raggedright Arriving \& Departing,
     Revised, Fit factor} & $\inflowvalue_i$  & $\inflowvalue_i$  & $\alpha\outflowvalue_i$ & $\checkmark$
     & \eqref{A:revised_modelH}
     \\
\end{tabular}
\caption[Different versions of the radiation model]{A summary of the different versions of the radiation model. The `Normalised?' column indicates if a model uses a normalisation that enforces the production constraint exactly  \eqref{e:radnormapp}, a cross indicates that \eqref{e:radgeneralapp} is used for that model. In each case we specify which of the site data values, ($P_i$ population, $I_i$ commuters arriving, $O_i$ commuters leaving) is used for the model site parameters (aspirations $\searchparam_i$, opportunities $\oppparam_i$, out flow $\flowparam_i$). See \tabref{t:param} for a summary of the notation.}
\label{t:radmodelsapp}
\end{table}

\subsection{The Populations Radiation model}

The `Populations' model (model A) is a standard radiation model \eqref{e:radgeneralapp} that sets all input parameters equal to the population ($m_i = n_i = t_i = P_i$). This gives us that
\begin{equation}
    \Fhat_{ij}
    =
    \popvalue_i \frac{\popvalue_i \popvalue_j}{(\popvalue_i + s_{ij})(\popvalue_i+ s_{ij} + \popvalue_j )} \, .
    \label{A:populations_modelA}
\end{equation}
Here the intervening opportunities measure $s_{ij}$ is the total population of sites lying closer to site $i$ than site $j$ (excluding site $i$ itself).

\subsection{The Departing Commuters Radiation model}

The `Departing Commuters' model (model B) is a standard radiation model  \eqref{e:radgeneralapp} defined as
\begin{equation}
        \Fhat_{ij}
    =
    \outflowvalue_i \frac{\outflowvalue_i \outflowvalue_j}{(\outflowvalue_i + s_{ij})(\outflowvalue_i + s_{ij} + \outflowvalue_j )} \, ,
    \label{A:departing_modelB}
\end{equation}
Here all input  parameters  to the number of commuters who depart from site $i$, $m_i = n_i = t_i = \outflowvalue_i$.
The intervening opportunities measure $s_{ij}$ in this model is the total number of commuters leaving all sites that are closer to site $i$ than site $j$ (excluding the commuters leaving site $i$ itself).
Note that in this model the total flow leaving site $i$ is \emph{not} equal to the number of commuters leaving site $i$, $\sum_{j \in \Tcal_i} \Fhat_{ij} \neq \outflowvalue_i$. This model has failed this normalisation criteria but in many cases this can be a small effect so this is not an unreasonable model to use.

\subsection{The Normalised Departing Commuters Radiation model}

The `Departing Commuters, Normalised' model (model C) is a normalised Radiation model \eqref{e:radnormapp} defined as
\begin{equation}
    \Fhat_{ij}
    =
    \left( \frac{\outflowvalue_i}{1 - \outflowvalue_i / N_c} \right)
     \frac{\outflowvalue_i \outflowvalue_j}{(\outflowvalue_i + s_{ij})(\outflowvalue_i + s_{ij} + \outflowvalue_j )} \, ,
    \qquad
    N_c = \sum_i \outflowvalue_i
    \label{A:departing_normed_modelC}
\end{equation}
Again all input parameters  to the number of commuters who depart from site $i$, $m_i = n_i = t_i = \outflowvalue_i$.
The intervening opportunities measure $s_{ij}$ is given in terms of the outputs of intervening sites, exactly as in the Departing Commuters Radiation model (model B) \eqref{A:departing_modelB}.
Unlike that model, this version is normalised properly so the total flow out of the model equals the associated data value exactly, $\sum_{j \in \Tcal_i} \Fhat_{ij} = \outflowvalue_i$ (see \eqref{e:radnormapp}).

\subsection{The Arriving \& Departing, Na\"{i}ve Split, Radiation Model}

The `Arriving \& Departing, Na\"{i}ve Split' radiation model (model D) is based on \eqref{e:radgeneralapp}, and defined as
\begin{equation}
    \Fhat_{ij}
    =
    \outflowvalue_i \frac{\outflowvalue_i \inflowvalue_j}{(\outflowvalue_i + s_{ij})(\outflowvalue_i + s_{ij} + \inflowvalue_j )}
     \, .
    \label{A:naive_modelD}
\end{equation}
Here both $m_i$ and $\flowparam_i$ (see \eqref{e:radgeneralapp}) are set equal to the number of commuters who depart from site $i$ so $m_i = \flowparam_i = \outflowvalue_i$.
We set the attractiveness model parameter, the number of opportunities at site $j$, to be equal to the total number of commuters found in the data to be arriving at site $j$, so $n_j=\inflowvalue_j$. This last identification then means that the intervening opportunities measure $s_{ij}$ has to be the cumulative number of commuters arriving at all sites closer to $i$ than $j$ (excluding site $i$), regardless of their origin.

\subsection{The Arriving \& Departing, Revised, Radiation Model}

The `Arriving \& Departing, Revised' model (model E) is a normalised Radiation model \eqref{e:radnormapp} defined as
\begin{equation}
    \Fhat_{ij}
    =
    \left( \frac{N_c}{N_c-\inflowvalue_i} \right)
    \outflowvalue_i \frac{\inflowvalue_i \inflowvalue_j}{(\inflowvalue_i + s_{ij})(\inflowvalue_i + s_{ij} + \inflowvalue_j )} \, ,
    \qquad
    N_c = \sum_i \inflowvalue_i \, .
    \label{A:revised_modelE}
\end{equation}
Here at each site $i$ the site attractiveness parameter $n_i$ and site aspiration parameter $m_i$ are both set equal to the number of commuters arriving at site $i$, $m_i=n_i=\inflowvalue_i$. The site repulsiveness parameter $\flowparam_i$ is set equal to the number of commuters leaving a site $\outflowvalue_i$ and the normalisation factor here ensures this is equal to the total flow predicted from the model, $\sum_{j} \Fhat_{ij} = \outflowvalue_i$.
The intervening opportunities measure $s_{ij}$ in this model is the total number of commuters arriving ($I_i$) at all sites that are closer to site $i$ than site $j$ (excluding the commuters arriving at site $i$ itself).

\subsection{The Populations, Additional Fitted Factor, Radiation model}\label{ass:departing_modelF}

The `Populations, Additional Fitted Factor' model (model F) is a standard radiation model \eqref{e:radgeneralapp} defined as
\begin{equation}
    \Fhat_{ij}
    =
    \alpha \popvalue_i \frac{\popvalue_i \popvalue_j}{(\popvalue_i + s_{ij})(\popvalue_i+ s_{ij} + \popvalue_j )} \, .
    \label{A:departing_modelF}
\end{equation}
Here we have $m_i=n_i = \popvalue_i$, but the flow parameter $\flowparam_i$ is set proportional to the total population $\flowparam_i = \alpha \popvalue_i$. This $\alpha$ is a single additional parameter found by optimising the fit to the data using a maximum likelihood estimation.
The intervening opportunities measure $s_{ij}$ is the total population of sites lying closer to site $i$ than site $j$ (excluding site $i$ itself).

\subsection{The Departing Commuters, Additional Fitted Factor, Radiation model}

The `Departing Commuters, Additional Fitted Factor' model (model G) is a standard radiation model \eqref{e:radgeneralapp} defined as
\begin{equation}
    \Fhat_{ij}
    =
    \alpha \outflowvalue_i
     \frac{\outflowvalue_i \outflowvalue_j}{(\outflowvalue_i + s_{ij})(\outflowvalue_i + s_{ij} + \outflowvalue_j )}
    \, .
    \label{A:departing_modelG}
\end{equation}
Here we have set $m_i=n_i = \outflowvalue_i$, the number of commuters who depart from site $i$, and $\flowparam_i = \alpha \outflowvalue_i$ where $\alpha$ is a single fitted parameter.  The intervening opportunities measure $s_{ij}$ is therefore the sum of all the outputs, $\outflowvalue_i$, of intervening sites.
Note that this model is not normalised, so $\sum_{j} \Fhat_{ij} \neq \alpha \outflowvalue_i$ and so in turn $\alpha=1$ is not to be expected even in `perfect' data generated from the model itself. In reality, the lack of accuracy in the model predictions is likely to ensure some $\alpha \neq 1$ will provide an optimal fit to the data.

\subsection{The Arriving \& Departing, Revised, Additional Fitted Factor, Radiation model}

The `Arriving \& Departing, Revised, Additional Fitted Factor' model (model H) is a normalised Radiation model \eqref{e:radnormapp}  defined as \begin{equation}
    \Fhat_{ij}
    =
    \alpha \outflowvalue_i
    \left( \frac{N_c}{N_c-\inflowvalue_i} \right)
    \frac{\inflowvalue_i \inflowvalue_j}{(\inflowvalue_i + s_{ij})(\inflowvalue_i + s_{ij} + \inflowvalue_j )} \, ,
    \qquad
    N_c = \sum_i \inflowvalue_i
    \label{A:revised_modelH}
\end{equation}
Here we have set $m_i=n_i = \inflowvalue_i$ and $\flowparam_i = \alpha \outflowvalue_i$. This model is a normalised radiation \eqref{e:radnormapp}. The intervening opportunities measure $s_{ij}$ is the total number of commuters arriving at all sites closer to $i$ than $j$.

The model is normalised so in principle we might expect $\alpha=1$. However we leave $\alpha$ as a single parameter to be found by optimising the fit to the data and so $\alpha \neq 1$ is likely given the inevitable imperfections in this simple model.


\section{Gravity Models}\label{a:gravity}

The Gravity model used Simini et al. \cite{SGMB12} was
\begin{equation}
 \Fhat_{ij}  =
 \theta(D-d_{ij})C_1 (P_i)^{\alpha_1} (P_j)^{\beta_1} (d_{ij})^{-\gamma_1}
 +
 \theta(d_{ij}-D)C_2 (P_i)^{\alpha_2} (P_j)^{\beta_2} (d_{ij})^{-\gamma_2}
 \, ,
 \label{A:gravitysimple}
\end{equation}
and similar with the exponential form for the deterrence function. The Heaviside theta functions split the model into short range and long range forms with a global model parameter $D$ determining the distance scale. This is a model with nine global parameters found through finding a best fit to the data. We have not used this form in our work.  We have instead worked with a production constrained gravity model with one parameter \eqref{e:radnormapp} as this matches the approximate production constraint and the single parameter used in the Radiation model of Simini et al. \cite{SGMB12} (the same as our Radiation model F of \eqref{A:departing_modelF}, see \tabref{t:radmodelsapp}).

\section{Applying common statistical techniques for comparing models}\label{a:common}

In this section, we discuss some of the commonly used statistical techniques for comparing models further, and apply these techniques to compare the models listed in \appref{A:radmodels} and the production constrained gravity model described by
\begin{equation}
    \Fhat_{ij} = \frac{ t_i n_j d_{ij}^{-\beta}}{\sum_k n_k d_{ik}^{-\beta}}
      \, .
      \quad (i \neq j)
    \label{e:gravitypcapp}
\end{equation}
It should be noted that any fitted parameters associated with the models have been calculated, as in the main text, through maximum likelihood estimation that utilises Poisson regression: only the techniques used for model comparison are the standard methods from the literature.

It should be noted that for most statistical methods, models whose predictions perfectly match real data can be identified even if the techniques are not technically theoretically justifiable. For example, a perfect model will always have $\mathrm{DSC} = R^2 = 1$. As such, we should expect some correlation between the results from these techniques and the results using our suggested, more accurate, methodology. However, in some cases, these techniques will give results that can be difficult to interpret -- or worse, will give wrong answers with no indication that the error is occurring.

\clearpage
\subsection{S{\o}rensen-Dice coefficient}

The S{\o}rensen-Dice coefficient can be applied to flows in the context of spatial modelling where it is sometimes referred to as the `common part of commuters' \cite{GLHB12,LHGD12}. It measures the overlap between the predicted and actual flow between each pair of sites, and is given by
\begin{equation}
    \mathrm{DSC} = \frac{\sum_{ij} \min(\Fhat_{ij}, F_{ij})}{\sum_{ij} F_{ij}}.
    \label{ae:DSC}
\end{equation}
It thus gives a result between $0$ and $1$, with a S{\o}rensen-Dice coefficient of $1$ representing a model with perfect predictions and $0$ a model with very weak predictions. The S{\o}rensen-Dice coefficient has been applied to spatial data in many cases, for instance \cite{GLHB12,LHGD12,WOETB15,KLGQ15,LBR16,GSSHSVSBR17,YZ19,LY20,YGZMWL20}.

One of the key limitations of the S{\o}rensen-Dice coefficient is that, in this form, it does not apply when the total predicted flow is not fixed to the total real flow (e.g.\ by a production constraint or weaker `total-flow' constraint). This is because without such a constraint, a model could predict infinite flows between all pairs of sites and obtain a perfect S{\o}rensen-Dice coefficient (equal to $1$). Perhaps most significantly, small percentage deviations in the predictions of large flows will have a more significant impact on the S{\o}rensen-Dice coefficient than large percentage deviations of small flows; the S{\o}rensen-Dice coefficient could more strongly penalise a model that consistently predicts flows to within 10\% of their actual value than a model which predicts its largest flows accurately but overestimates tiny flows by several orders of magnitude. Thus, models which make better overall predictions, when judged by a more rigorous Poisson regression analysis, can have a worse S{\o}rensen-Dice coefficient, and, additionally, differences between the S{\o}rensen-Dice coefficients of different models can be difficult to interpret.

Variations on the form of the DSC can resolve the first issue, by multiplying the numerator by two and adding $\sum_{ij} \Fhat_{ij}$ to the denominator \cite{MSJB13,YHEG14}, perhaps applied only to links with non-zero values, the `common part of links' measure used in \cite{LBR16}. However, this amendment does not resolve the other limitations.

\begin{figure}[htb!]
	\centering
    \includegraphics[width=\textwidth]{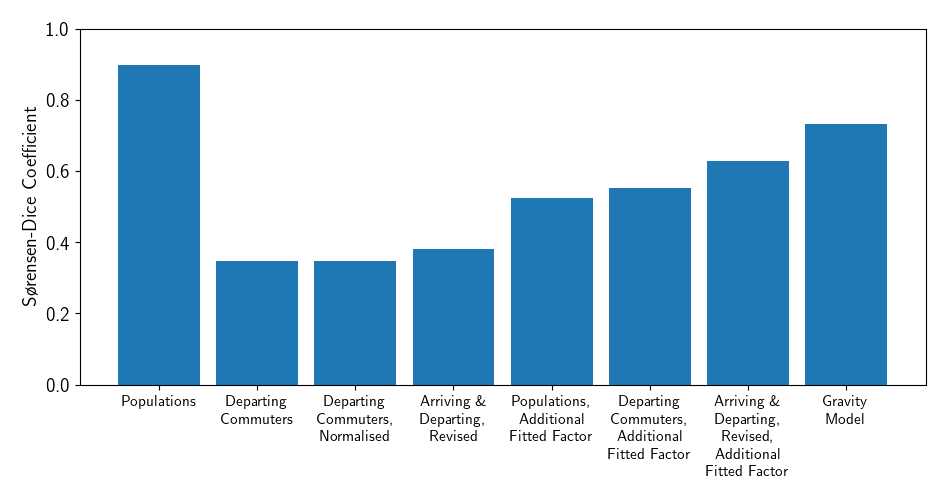}
	\caption{S{\o}rensen-Dice coefficient values \eqref{ae:DSC} for radiation models A, B, C, E, F, G, H (from left to right) described in \tabref{t:radmodelsapp}, alongside the production constrained gravity model of \eqref{e:gravitypcapp}. Higher values represent better models. These data are from the US Census 2000 \cite{US_census_2000}.}
	\label{fig:DSC}
\end{figure}

S{\o}rensen-Dice coefficients for the models we have analysed appear in \figref{fig:DSC}. The results would indicate that model A (`Populations') performs best by this measure, however since this model lacks a total flow constraint, the associated S{\o}rensen-Dice coefficient is not meaningful. While some of the S{\o}rensen-Dice coefficient results follow the same pattern as those found from our maximum likelihood methods, with model B (`Departing Commuters') being the weakest of the models shown (excluding model A), and the gravity model giving the best result, some of the results are dissimilar. For example, by our methods model E is shown to have greater predictive power than model F, which is not captured by the S{\o}rensen-Dice coefficient.

\clearpage
\subsection{Kolmogorov-Smirnov Test}

The Kolmogorov-Smirnov test is defined  \cite{steinskog_cautionary_2007}:
\begin{equation}
    K = \sup|\Fhat_{ij} - F_{ij}|.
    \label{ae:KS}
\end{equation}
This is a nonparametric test of the equality of two continuous valued functions, which in our case are the data and the model prediction of the flow between each pair of sites.
The test may return values between $0$ and $+\infty$, with smaller values representing better models. While for pairs of sites with high flows, the integer value may be reasonably approximated by a continuous function, the integer valued nature of the data will be a concern for the majority of site pairs where flows are low.

However our main concern is that the Kolmogorov-Smirnov test requires its two input functions to be independent.
When the Kolmogorov-Smirnov test is used in spatial modelling \cite{kang_generalized_2015}, the test is usually applied to a model whose parameters have been estimated by fitting to the same data so not the data and model functions are not independent \cite{steinskog_cautionary_2007}. When the conditions for the validity of this test are not met, the results can appear absurd. For example, when comparing two models, one of which is very accurate, and one of which is not, but both of which have a single large outlier, the Kolmogorov-Smirnov test will return large values for both models.

\begin{figure}[htb!]
	\centering
    \includegraphics[width=\textwidth]{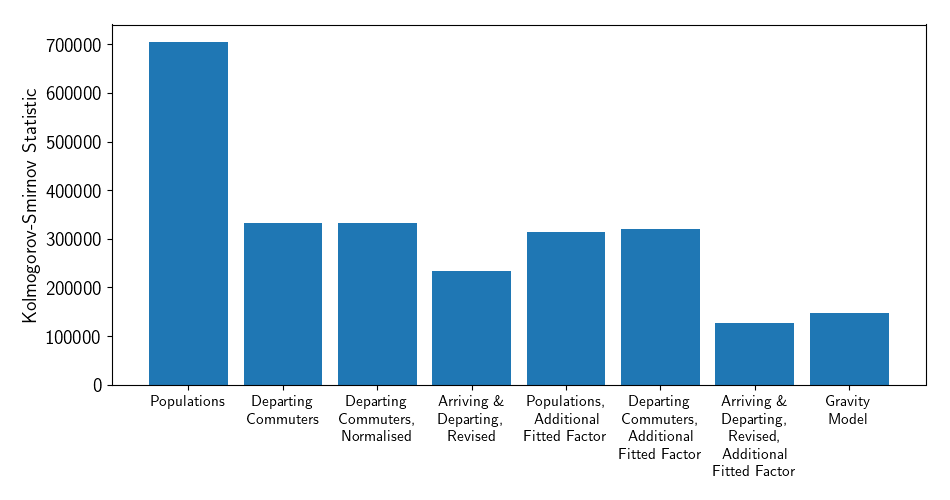}
	\caption{Kolmogorov-Smirnov values \eqref{ae:KS} for radiation models A, B, C, E, F, G, H (from left to right) described in \tabref{t:radmodelsapp}, alongside the production constrained gravity model of \eqref{e:gravitypcapp}. Lower values represent better models. These data are from the US Census 2000  \cite{US_census_2000}.}
	\label{fig:KS}
\end{figure}

The results found by applying the Kolmogorov-Smirnov test to our models are shown in \figref{fig:KS}. As with the S{\o}rensen-Dice coefficient, there is some correlation between the ranking of models obtained using the Kolmogorov-Smirnov statistic and the ranking of models obtained using our more rigourous analysis above. For example, model A (`Populations') is shown to be the worst of the models depicted in this figure. However, there are also notable differences -- the gravity model is, according to the Kolmogorov-Smirnov test, worse than model H (`Arriving \& Departing, Revised, Additional Fitted Factor).

\clearpage
\subsection{The Coefficient of Determination, $R^{2}$}

The coefficient of determination $R^2$ is one of a family of measures which assume that the error distribution $p(F_{ij}| \Fhat_{ij})$ is Gaussian for any $i$, $j$. However, it is very common that real data sets feature no negative flows and very many small flows, meaning that the central limit theorem does not apply and the distributions of flows between any pair of sites cannot be assumed to be Gaussian. As such $R^2$ should not be assumed to be a theoretically valid measure when used to analyse predictions from spatial interaction models.

\begin{figure}[htb!]
	\centering
    \includegraphics[width=\textwidth]{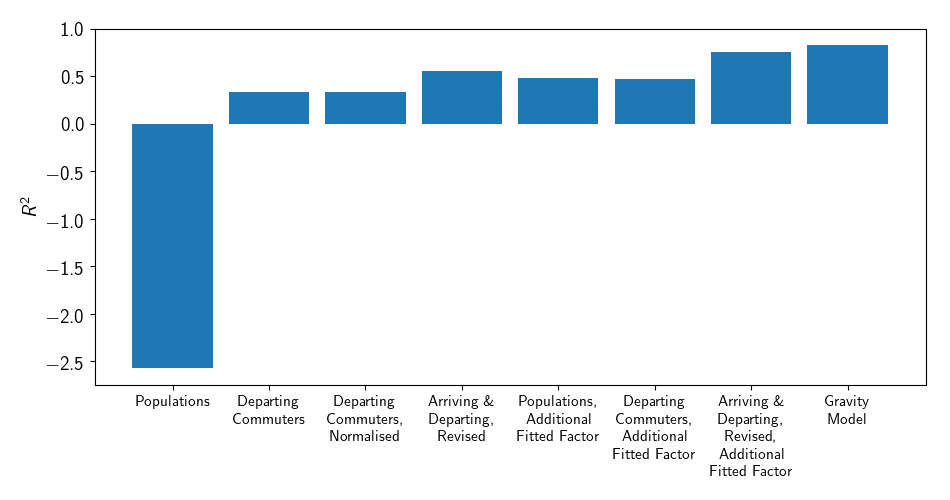}
	\caption{$R^2$ values \eqref{e:R2} for radiation models A, B, C, E, F, G, H (from left to right) described in \tabref{t:radmodelsapp}, alongside the production constrained gravity model of \eqref{e:gravitypcapp}. Lower values represent better models. These data are from the US Census 2000  \cite{US_census_2000}.}
	\label{fig:R2}
\end{figure}

The coefficient of determination is given by
\begin{equation}
    R^2 = 1 - \frac{\sum_{ij} (F_{ij} - \hat F_{ij})^2} {\sum_{ij} (F_{ij} - \bar F)^2},
    \label{e:R2}
\end{equation} where $\bar F$ is the mean value of the data set. $R^2$ values can range from $-\infty$ to $1$, with values closer to $1$ representing better models. The $R^2$ values obtained for our models are shown in \figref{fig:R2}.
This measure has been used to assess the goodness-of-fit of models to data in spatial contexts \cite{MSJB13,HJ16,BPTC16}.

The first thing to note is that whilst, in general, models shown to be better in our more rigourous analysis above are better using $R^2$ values, there are key differences in the rankings. For example, model F (`Populations, Additional Fitted Factor') and model G (`Departing Commuters, Additional Fitted Factor') are, according to these $R^2$ values, more successful than  model E (`Arriving \& Departing, Revised').

Most notably, model A (`Populations') has a negative $R^2$. Usually, this would mean that the model being tested is worse than the null hypothesis --- for $R^2$, the null hypothesis is a horizontal line. However, in the (theoretically unjustifiable) way in which the $R^2$ has been used here, we cannot easily interpret the negative $R^2$ in this way. This is illustrative of the general problem of using statistical methods that are not theoretically justifiable in the case of spatial interaction modelling --- it is very difficult to legitimately interpret the values obtained.

There are other similar measures based on squared differences, so typically motivated by Gaussian statistics, and these include mean squared errors \cite{CPGB18} and Pearson correlation coefficients \cite{LZKCJ15,GETGBMW20}. We do not pursue them here.

\clearpage
\section{Data}\label{a:data}

The distribution of commuter flows in the US Census 2000  \cite{US_census_2000} data is shown in \tabref{tab:flownumber}.

\begin{table}[htb]
 \centering
 \begin{tabular}{r|r}
       Flow & Number \\ \hline
        All & 9665881 \\
       $>0$ & 164764 \\
      $>10$ & 77432 \\
     $>100$ & 21237 \\
   $>1,000$ & 7058 \\
  $>10,000$ & 1814 \\
 $>100,000$ & 212 \\
 \end{tabular}
 \caption[Number of Flows]{The number of county-county pairs with flows equal to or greater than the flow minimum given.  Data for county-county commuter numbers is as given in the US Census 2000  \cite{US_census_2000}.}
 \label{tab:flownumber}
\end{table}

\end{document}